\def\mbf{\mathbf }
\def\k{{\mathbf k}\ }
\def\n{{\noindent}}
\def\q{{\mathbf q}}

\def\g{{\mathbf g}}
\def\D{{\mathbf D}}
\def\A{{\mathbf A}}
\def\B{{\mathbf B}}
\def\F{{\mathbf F}}
\def\S{{\mathbf S}}
\def\wt{\widetilde }
\def\n{\noindent}

\def\be{\begin{equation}}
\def\ee{\end{equation}}

\def\d{{\rm d}}

\def\P{$P$}

\def\eq{\enskip =\enskip}

\def\mns{\enskip -\enskip}

  \def\ket{\vert \vert  \{ \emptyset \} \rangle}
  \def\ket2{\vert \vert \otimes \{ R \} \rangle}

\def\.#1{\mathaccent 95#1}
\def\^#1{\mathaccent 94 #1}
\def\~#1{\mathaccent "7E #1}

\def\eq{\enskip =\enskip}

\def\mns{\enskip -\enskip}

  \def\ket{\vert \vert  \{ \emptyset \} \rangle}
  \def\ket2{\vert \vert \otimes \{ R \} \rangle}
  
\def\k{{\bf k}}

\def\r{\underline{r} }

\documentclass[12pt,epsfig]{iopart}
\usepackage{graphicx}% Include figure files
\begin{document}
\jl{3}
\title[Phonons in disordered alloys : a multiple scattering approach]{Phonons in disordered
alloys : a multiple scattering approach}
\author{Aftab Alam\footnote{corresponding author, email : alam@bose.res.in} and  
Abhijit Mookerjee\footnote[1]{email : abhijit@bose.res.in}}
\address{Satyendra Nath Bose National Centre for Basic Sciences, JD Block, Sector III, Salt Lake City, Kolkata 700098, India}

\begin{abstract}
In this paper we shall discuss the effect of disorder induced configuration
fluctuations on single particle and two-particle phonon Green functions in
substitutional random binary alloys. 
The randomness of the system will be dealt within the augmented space theorem, introduced by one of us \cite{am1}. This will be combined with a generalized 
Edwards-Langer diagrammatic technique to extract various useful results in the form of mathematical expressions. The purpose of present manuscript will be
 to describe the scattering diagrams in full detail. This will form the
basis of the calculational algorithms which we had used in an earlier
 paper on NiPd and NiPt alloys \cite{am,am2}. We shall show the structure of all
 possible scattering diagrams up to the 4-th order and subsequently  
illustrate how to obtain Dyson's equation from a re-summation of
the  diagrammatic series. We shall also study how disorder scattering affects
two-particle Green functions associated with thermal response. 
 We shall show that disorder scattering renormalizes both the 
phonon propagators as well as the heat currents. 
These renormalized heat currents will be shown to be related to the self-energy
 of the propagators. We shall also study  a different class of scattering diagrams which are not related to the self-energy but rather to the vertex corrections.  
\end{abstract} 

\section{Introduction}

In an earlier communication \cite{am} we had discussed the calculation of the 
configuration averaged thermal conductivity and diffusivities of disordered NiPt
 and NiPd alloys. We had argued that 
lattice thermal conductivity yields valuable information 
about the interactions of thermal excitations with composition fluctuations
on the crystal lattice. 
 We had also argued that although the theory of lattice thermal conductivity for perfect crystals and ordered compounds has been developed,  the same is not true for disordered alloys. 
The presence of disorder results in scattering that not only depends on the 
impurity concentration but also crucially on both the relative masses and 
size difference between the constituent atoms as well as the difference
between their dynamical matrices. For large mass and/or dynamical matrix differences, 
the effect of disorder can be quite
unusual. That earlier communication focused on the results of numerical calculations
on NiPt and NiPd and their comparison with available experimental data. In this communication we shall present a detailed analysis of the effects of disorder scattering on the thermal
conductivity. We shall use the Edwards-Langer scattering formalism \cite{ed,la} coupled with the augmented space technique introduced by 
our group \cite{tf} to
analyze  the disorder effect. This analysis will form the basis for justification of the calculational algorithms used by us in our earlier paper.

\section{Effect of disorder on the single particle Green function : the Dyson equation}

The augmented space formalism (ASF) for carrying out configuration averaging of physical properties of disordered systems has been described in detail in several earlier papers (see \cite{am1},\cite{am2},\cite{tf}). We shall, for the sake
of completeness,  describe only those features which will be necessary for the implementation of our ideas in this
communication.

Let $f(n_R)$ be a function of a binary random variable $n_R$ , whose binary probability density is given by :

\[ Pr(n_R) \ =\  x_A\ \delta(n_R)\ +\  x_B\ \delta(n_R-1) \]

Such random variables are useful for describing randomly substitutional binary alloys. These are the kind of disordered systems we shall be interested in for this communication. 
The ASF now prescribes that we write this probability density, which is a 
positive definite integrable function as the resolvent of an operator whose 
spectrum consists of the random values taken by it (in this case 0 and 1) : 

\begin{equation}
 Pr(n_R) \ =\  -\frac{1}{\pi}\ \Im m \langle\uparrow_R \vert (n_R I - N_R)^{-1}\vert\uparrow_R\rangle
\label{prob} 
\end{equation}
The {\sl configuration space} of $n_R$ is of rank two and spanned by the {\sl states} $\vert 0\rangle$ and $\vert 1\rangle$.
The operator $N_R$ acts on  this space.  
$\vert\uparrow_R\rangle =\sqrt{x_A}\vert 0\rangle +\sqrt{x_B}\vert 1\rangle $
is called the {\sl reference state}. Its orthogonal counterpart is 
$\vert\downarrow_R\rangle =\sqrt{x_B}\vert 0\rangle -\sqrt{x_A}\vert 1\rangle $. The representation of $N_R$ in this new basis is :

\[ N_R \ =\  \left( \begin{array}{cc}
                   x_A & \sqrt{x_A x_B} \\
               \sqrt{x_A x_B} & x_B 
                   \end{array} \right) \]

The ASF now proceeds as follows :

\begin{eqnarray}
\fl \ll f(n_R)\gg\   =\  \int_{-\infty}^{\infty}\  f(n_R)Pr(n_R) dn_R 
 \ =\ -\frac{1}{\pi}\Im m \int_{-\infty}^{\infty}\ f(n_R)\langle\uparrow_R\vert (n_RI-N_R)^{-1}
 \vert\uparrow_R\rangle dn_{R}\nonumber\\
= \  -\frac{1}{\pi}\Im m \sum_{\lambda=0,1}\sum_{\lambda '=0,1} \int_{-\infty}^{\infty}\ f(n_R)
\langle\uparrow_R\vert\lambda\rangle
 \langle\lambda\vert (n_RI-N_R)^{-1}\vert\lambda '\rangle\langle\lambda'\vert\uparrow_R\rangle dn_{R} \nonumber\\
= \ \sum_{\lambda =0,1}  \langle\uparrow_R\vert\lambda\rangle f(\lambda)\langle \lambda\vert\uparrow_R\rangle
\ =\  \ \langle\uparrow_R\vert \tilde{\mathbf f}\vert \uparrow_R\rangle
\end{eqnarray}

Here $\tilde{\bf f}$ is an operator built out of $f(n_R)$ by simply replacing the variable $n_R$ by the associated operator $N_R$. The
above expression shows that the average is obtained by taking the matrix element of this operator between the {\sl reference state}
$\vert\uparrow_R\rangle$. The full Augmented Space Theorem is a generalization of this for functions of many independent random 
variables $\{n_R\}$.

The theory of phonons consists of solving a secular equation of the form~: 
 \[ ({\bf M}w^{2} - {\bf D})\ {\bf u}(R,w) = 0 \]  where $u_{\alpha}(R,w)$ is the Fourier transform of $u_{\alpha}(R,t)$, the displacement of an atom from its equilibrium position $R$ on the lattice, in the direction ${\alpha} $ at time $t$. {\bf M} is the {\it mass operator}, diagonal in real-space
 and {\bf D} is the {\it dynamical matrix operator} whose tight-binding representations are :
\begin{eqnarray}
{\bf M} &=& \sum_{R}  m_{R}\ {\delta}_{\alpha \beta} \ P_R\nonumber\\
{\bf D} &=&  \sum_{R} \left\{-\sum_{R^{\prime} \ne R}\Phi_{RR^{\prime}}^{\alpha \beta}\right\}\ P_{R} + \sum_{R}\sum_{R^{\prime} \ne R} \Phi_{RR^{\prime}}^{\alpha \beta}\ T_{RR^{\prime}}\nonumber\\
\end{eqnarray}
\n where the sum rule has been incorporated in the first term of the equation involving $\bf D$. \\
Here $P_R$ is the projection operator\ $\vert R\rangle\langle R\vert$\ \ and $T_{RR'}$ is the transfer operator\ $\vert R\rangle\langle R'\vert$\ \ in the real space ${\cal H}$ spanned by the tight-binding basis $\{\vert R\rangle\}$.
 $R,R^{\prime}$ specify the lattice sites and $\alpha $,$ \beta $ the Cartesian directions. $m_{R}$ is the mass of an atom occupying the position $R$ and $\Phi_{RR^{\prime}}^{\alpha \beta}$ is the force constant tensor. 

\noindent We shall be interested in calculating the displacement-displacement Green matrix ${\mbf G}
(R,R',w^2)$ :

\[ {\mbf G}(R,R',w^2) = \langle R | \left({\bf{M}}w^{2}-{\bf{D}}\right)^{-1} | R' \rangle \]

Let us now consider a binary alloy $ A_{x}B_{y} $ consisting of two kinds of atoms A and B of masses
 $m_A$ and  $m_B$ randomly occupying each lattice sites. We wish to calculate the configuration-averaged
 Green matrix $\ll {\bf G}({R,R'},w^2)\gg$. We shall use the augmented space formalism to do so indicating the main operational results here.  For further details we  refer the reader to the above monograph \cite{tf}.
The first operation is to represent the random parts of the secular equation in terms of a random
set of local variables $\{ n_R\}$ which are 1 if the site $R$ is occupied by an  A  atom and 0
if it is occupied by B. 

In terms of these, the mass operator can be written as :
\begin{equation} 
{\bf M}\ =\ \sum_{R}  \left[\rule{0mm}{4mm} m_{B}\ +\ n_{R}\ (\delta m) \right] \delta_{\alpha \beta} \ P_R \quad ;\quad \delta m=m_A-m_B
\end{equation}

\noindent According to the augmented space theorem, in order to obtain the configuration-average we simply replace the random variables $n_R$ by the
corresponding operators $N_R$ associated with its probability density given by equation (\ref{prob}), and take the matrix element of
the resulting operator between the {\sl reference states}. For a full
mathematical proof the reader is referred to \cite{tf}.

\[
 n_{R}\longrightarrow N_{R} \ =\  
x\ \tilde{I}\ +\ (y-x)\  p_{R}^{\downarrow} + \sqrt{xy}\ {\cal T}^{\uparrow \downarrow}_R  
\]

 Similarly the random off-diagonal force constants $\Phi_{RR^{\prime}}^{\alpha \beta}$ between the sites $R$ and $R^{\prime}$ can be written as~:

\begin{eqnarray}
\Phi_{RR^{\prime}}^{\alpha \beta} &=& \Phi_{AA}^{\alpha \beta} n_{R} n_{R^{\prime}} + \Phi_{BB}^{\alpha \beta} (1-n_{R}) (1-n_{R^{\prime}})\ +\  \Phi_{AB}^{\alpha \beta} \left[\rule{0mm}{4mm}\ n_{R}(1-n_{R^{\prime}}) + n_{R^{\prime}}(1-n_{R})\ \right]\nonumber\\
\phantom{x} \nonumber\\
&=&  \Phi_{BB}^{\alpha \beta}\ +\ \left(\rule{0mm}{4mm}\Phi_{AA}^{\alpha \beta} + \Phi_{BB}^{\alpha \beta} - 2 \Phi_{AB}^{\alpha \beta}\right)\ n_{R} n_{R^{\prime}}\ 
 +\  \left(\rule{0mm}{4mm}\Phi_{AB}^{\alpha \beta} - \Phi_{BB}^{\alpha \beta}\rule{0mm}{4mm}\right)\ (n_{R} + n_{R^{\prime}}) \nonumber \\
\end{eqnarray}

The augmented space theorem \cite{am1} yields the configuration averaged Green function\cite{am2} as
\begin{equation}
\ll {\mbf G}(R,R',w^2)\gg = \langle\{\emptyset\}\otimes R\vert \left({\mbf g}^{-1}- \widetilde{\mbf D}_1\right)^{-1}\vert\{\emptyset\}\otimes R'\rangle,
\label{eq1}
\end{equation}
\n where the VCA green matrix ${\bf g}$ is given by
\begin{equation}
{\mbf g} =  ( \ll \widetilde{\mbf M}\gg \omega^{2}  - \ll\widetilde{\bf D}\gg)^{-1} ,\nonumber
\label{eq2}
\end{equation}
\n with the Mass operator ${\widetilde\mbf M}$ and the Dynamical matrix operator ${\widetilde\mbf D}$ in the augmented space has the form
\begin{eqnarray}
\widetilde{\bf M}&=& \A({\mbf m} )\ \widetilde{I}\otimes  I + \B({\mbf m} )\ \sum_{R} p_{R}^\downarrow \otimes P_{R} + \F({\mathbf m} )\ \sum_{R}\  {\cal T}^{\uparrow\downarrow}_R \otimes P_{R},\nonumber\\ 
\phantom{\widetilde{\bf M}} &=& \ll \widetilde{\mbf M}\gg\  +\  \widetilde{\mbf M}^\prime,\nonumber
\end{eqnarray}
{\n where}
\[   \begin{array}{ll}

\A({\mathbf X}) = \ll {\mbf X} \gg\ =(x {\mathbf X}_A+y {\mathbf X}_B), \\
\B({\mathbf X}) = (y-x)\ ({\mathbf X}_A-{\mathbf X}_B), \\
\F({\mathbf X}) = \sqrt{xy} \ ({\mathbf X}_A-{\mathbf X}_B).  \end{array}  \]

x,y being the concentration of the components A and B of the alloy $A_xB_y$. $p_{R}^\downarrow$ and ${\cal T}^{\uparrow\downarrow}_R$ are the projection and transfer operators in the configuration space describing the statistical behaviour of the system.

\begin{eqnarray}
\fl\ll{\wt \D}\gg = -\sum_{R}\sum_{R'\ne R} \ll \Phi_{RR'}\gg \widetilde {I}\otimes P_R  + \sum_{R}\  \sum_{R'\ne R}  \ll\Phi_{RR'}\gg\widetilde {I}\otimes T_{RR^{\prime}}
\label{eq3}
\end{eqnarray}

$P_R$ and $T_{RR^{\prime}}$ are the projection and transfer operators in the real Hilbert space.

\begin{eqnarray}
\ll \Phi_{RR^{\prime}}\gg=x^{2}\ \Phi_{AA}+ y^{2}\ \Phi_{BB}+ 2xy\ \Phi_{AB}.\nonumber
\end{eqnarray}

$\Phi's$ in the right hand side are  the force constant tensor between different combinations of atoms.
\n Also

\begin{eqnarray*}
\fl\widetilde{\mathbf D}_1 = \sum_R \  \left\{\rule{0mm}{5mm} {- \mbf \Upsilon}_R -\sum_{R'\ne R} {\mathbf\Psi}_{RR'}\right\}\ \otimes P_R + 
\sum_{R}\sum_{R'\ne R}\  {\mathbf\Psi}_{RR'}\otimes T_{RR'}
\end{eqnarray*}
\n with
\begin{eqnarray}
{\mbf\Upsilon}_R & = & \B({\mbf m})\ \omega^2\ p^\downarrow_R \ +\ \F({\mbf m})\ \omega^2\ {\cal T}^{\uparrow\downarrow}_R
\nonumber\\
\phantom{x}\nonumber\\
 {\mathbf\Psi}_{RR'}& = & {\mathbf D}^{(1)}_{RR'}\  \left(p^\downarrow_R +  p^\downarrow_{R'}\right)
+ {\mathbf D}^{(2)}_{RR'}\ \left({\cal T}^{\uparrow\downarrow}_R + {\cal T}^{\uparrow\downarrow}_{R'}\right) +
 {\mathbf D}^{(3)}_{RR'}\  p^\downarrow_R\  p^\downarrow_{R'}+ \nonumber\\
& & {\mathbf D}^{(4)}_{RR'}\ \left( p^\downarrow_R\ {\cal T}^{\uparrow\downarrow}_{R'} + 
 {\cal T}^{\uparrow\downarrow}_{R}\ p^\downarrow_{R'}\right)
 +\ {\mathbf D}^{(5)}_{RR'}\ {\cal T}^{\uparrow\downarrow}_R
\ {\cal T}^{\uparrow\downarrow}_{R'},
\label{eq4}
\end{eqnarray}
\n where
\begin{eqnarray*}
{\mathbf D}^{(1)}=(y-x)\ \Phi_{(1)}\hspace{1in}
{\mathbf D}^{(2)}=\sqrt{xy}\ \Phi_{(1)},\\
{\mathbf D}^{(3)}={(y-x)^{2}}\ \Phi_{(2)}\hspace{1in}
{\mathbf D}^{(4)}=\sqrt{xy}\ (y-x)\ \Phi_{(2)},\\
{\mathbf D}^{(5)}= xy\ \Phi_{(2)}.\\
\end{eqnarray*}
\n and
\begin{eqnarray*}
\Phi_{(1)} &=& x\ \Phi_{AA} - y\ \Phi_{BB} + (y-x) \Phi_{AB},\\
\Phi_{(2)} &=& \Phi_{AA} + \Phi_{BB} - 2 \Phi_{AB}.
\end{eqnarray*}

\begin{figure}[h]
\centering
\includegraphics[width=10cm,height=8.5cm]{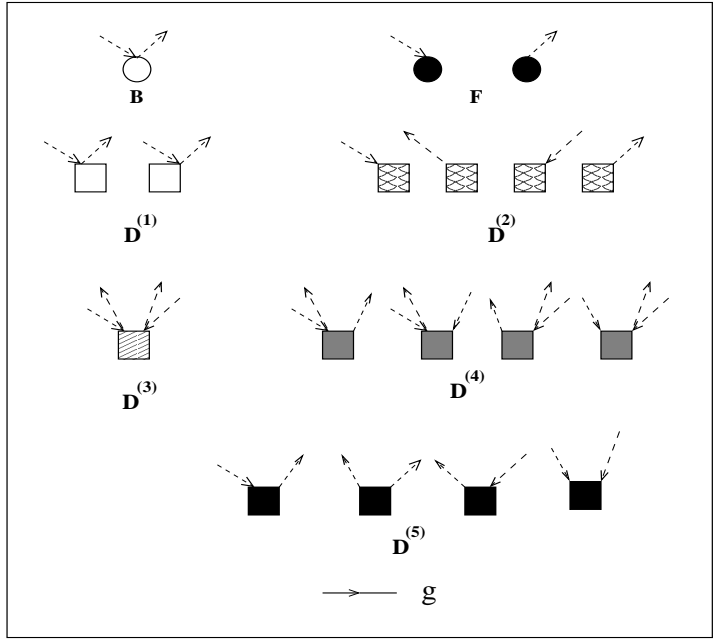}
\caption{The scattering vertices for the averaged Green function}
\label{fig1}
\end{figure}

Physical interpretation of the vertices is transparent if we consider the $\vert\uparrow_R\rangle$ state to be the {\sl ground state} or a state with no configuration fluctuations
and $\vert\downarrow_R\rangle$ to be the state with one configuration fluctuation at the site $R$. The operator $p^\downarrow_R$ then measures the number of fluctuations (0 or 1) and
${\cal T}^{\uparrow\downarrow}_R$ either creates or destroys a fluctuation at the site R.
From equation (\ref{eq3}) then we can interpret the following :

The scattering vertices {\bf B},{\bf F} and {\bf D}$^{(1)}$-{\bf D}$^{(5)}$ can be represented as shown
in Figure \ref{fig1}. The vertex {\bf F} is diagonal in real space and causes a
configuration fluctuation at a site due to disorder in the mass. Similarly, the
vertex {\bf B}, also diagonal in real space,  counts the number of fluctuations at a given site.
The remaining vertices labelled {\bf D} are all off-diagonal in real space. The vertex
{\bf D}$^{(1)}$ counts the number of fluctuations at one or the other of the sites associated with the vertex. {\bf D}$^{(2)}$ causes a configuration fluctuation, again at
one of the sites associated with the vertex. {\bf D}$^{(3)}$ counts the number of
fluctuations at both the sites associated with the vertex. {\bf D}$^{(4)}$ counts the
number of fluctuations at one site  and causes a fluctuation at the other. Finally, {\bf D}$^{(5)}$  causes a fluctuation at both the sites associated with the vertex.

Equation (\ref{eq1}) can also be expressed as
\begin{eqnarray}
\fl \ll {\mbf G}(R,R',w^2)\gg 
 =  \langle \{\emptyset\}\otimes R \vert \left(\rule{0mm}{5mm} \g + \g\ \widetilde{\mbf D}_1\ \g
+ \g\ \widetilde{\mbf D}_1\ \g\ \widetilde{\mbf D}_1\ \g + \ldots \right) \vert \{\emptyset\}\otimes R'\rangle \nonumber\\
\label{eq5}
\end{eqnarray}
The first term in Equation (\ref{eq5}) gives :
\begin{eqnarray}
\langle \{\emptyset\}\otimes R \vert\ \g\  \vert  \{\emptyset\}\otimes R' \rangle \ =\ {\mbf g} (R,R',w^{2})\nonumber 
\end{eqnarray}
In scattering diagram formalism, we shall associate a propagator represented by a horizontal arrow for each factor $\g$.

\n The second term in equation ({\ref{eq5}}) yields zero since $\langle \{\emptyset\}\otimes R \vert \widetilde{\mbf D}_1 \vert \{\emptyset\}\otimes R' \rangle = 0 $. 

The third term gives~:
\begin{eqnarray*}
\hspace{-2.0cm}\langle\{\emptyset\}\otimes R\vert\g\ \widetilde{\mbf D}_1\ \g\ \widetilde{\mbf D}_1\ \g  \vert\{\emptyset\}\otimes R^{\prime}\rangle =\\
 \sum_{S_{1}S_{2}}\sum_{S_{3} S_{4}}\sum_{\{\cal C\}}\sum_{\{\cal C'\}}\ \langle \{\emptyset\}\otimes R\vert\ \g\ \vert \{\emptyset\}\otimes S_{1}\rangle \langle \{\emptyset\}\otimes S_{1}\vert\ \wt{\mbf D}_1\ \vert \{{\cal C}\}\otimes S^{\prime\prime}\rangle\\ \langle \{{\cal C}\}\otimes S_{2}\vert\ \g\  \vert \{{\cal C}'\}\otimes S_{3}\rangle \langle \{{\cal C}'\}\otimes S_{3}\vert\ \wt{\mbf D}_1\ \vert \{\emptyset\}\otimes S_{4} \rangle\ \langle \{\emptyset\}\otimes S_{4}\vert\ \g\ \vert \{\emptyset\}\otimes R'\rangle 
\end{eqnarray*}
\mbox {A little more algebra yields the following contribution :}

\begin{eqnarray}
\hspace{-2.3cm}\langle\{\emptyset\}\otimes R\vert\g\ \widetilde{\mbf D}_1\ \g\ \widetilde{\mbf D}_1\ \g  \vert\{\emptyset\}\otimes R^{\prime}\rangle =\nonumber\\
\hspace{-1.5cm} \sum_{S_1S_2}\ \g(R,S_1,w^2)\  (\F w^2)\  \g(S_1,S_2,w^2)\ \delta(S_1-S_2)\ (\F w^2)\ \g(S_2,R',w^2) +  \nonumber\\
\hspace{-1.5cm}\sum_{S_1}\sum_{S_3S_4}\ \g(R,S_1,w^2)\ (\F w^2)\  \g(S_1,S_3,w^2)\ \delta(S_1-S_3)\ \D^{(2)}_{S_3S_4}\ \g(S_4,R',w^2) + \nonumber\\
\hspace{-1.5cm} \sum_{S_1}\sum_{S_3S_4}\ \g(R,S_1,w^2)\ (\F w^2)\  \g(S_1,S_3,w^2)\ \delta(S_1-S_4)\ \D^{(2)}_{S_3S_4}\ \g(S_4,R',w^2) + \nonumber\\
\hspace{-1.5cm} \sum_{S_1}\sum_{S_2S_4}\ \g(R,S_1,w^2)\  \D^{(2)}_{S_1S_2}\ \g(S_2,S_4,w^2)\ \delta(S_1-S_4)\ (\F w^2)\ \g(S_4,R',w^2) + \nonumber\\
\hspace{-1.5cm} \sum_{S_1}\sum_{S_2S_4}\ \g(R,S_1,w^2)\  \D^{(2)}_{S_1S_2}\ \g(S_2,S_4,w^2)\ \delta(S_2-S_4)\ (\F w^2)\ \g(S_4,R',w^2) + \nonumber\\
\hspace{-1.5cm} \sum_{S_1S_2}\sum_{S_3S_4}\ \g(R,S_1,w^2)\  \D^{(2)}_{S_1S_2}\ \g(S_2,S_3,w^2)\ \delta(S_1-S_3)\ \D^{(2)}_{S_3S_4}\ \g(S_4,R',w^2) + \nonumber\\
\hspace{-1.5cm} \sum_{S_1S_2}\sum_{S_3S_4}\ \g(R,S_1,w^2)\  \D^{(2)}_{S_1S_2}\ \g(S_2,S_3,w^2)  \delta(S_1-S_4)\ \D^{(2)}_{S_3S_4}\ \g(S_4,R',w^2) + \nonumber\\
\hspace{-1.5cm} \sum_{S_1S_2}\sum_{S_3S_4}\ \g(R,S_1,w^2)\  \D^{(2)}_{S_1S_2}\ \g(S_2,S_3,w^2)\ \delta(S_2-S_3)\ \D^{(2)}_{S_3S_4}\ \g(S_4,R',w^2) + \nonumber\\
\hspace{-1.5cm} \sum_{S_1S_2}\sum_{S_3S_4}\ \g(R,S_1,w^2)\  \D^{(2)}_{S_1S_2}\ \g(S_2,S_3,w^2)\ \delta(S_2-S_4)\ \D^{(2)}_{S_3S_4}\ \g(S_4,R',w^2) + \nonumber\\
\hspace{-2.3cm} \sum_{S_1S_2}\sum_{S_3S_4}\ \g(R,S_1,w^2)\  \D^{(5)}_{S_1S_2}\  \g(S_2,S_3,w^2)\ \delta(S_2-S_4)\delta(S_1-S_3)\ \D^{(5)}_{S_3S_4}\ \g(S_4,R',w^2) + \nonumber\\
\hspace{-2.3cm} \sum_{S_1S_2}\sum_{S_3S_4}\ \g(R,S_1,w^2)\ \D^{(5)}_{S_1S_2}\ \g(S_2,S_3,w^2)\ \delta(S_1-S_4)\delta(S_2-S_3)\ \D^{(5)}_{S_3S_4}\ \g(S_4,R',w^2) \nonumber\\
\label{eq6}
\end{eqnarray}

Referring to the above equation, we shall now build up the scattering diagrams. We
have already associated  scattering vertices with the terms in $\wt \D_{1}$. These are the seven different types of scattering vertices as shown in Fig. \ref{fig1}. The dashed lines with arrow are associated with the delta function.

\begin{figure}
\centering
\includegraphics[height=9cm,width=10cm]{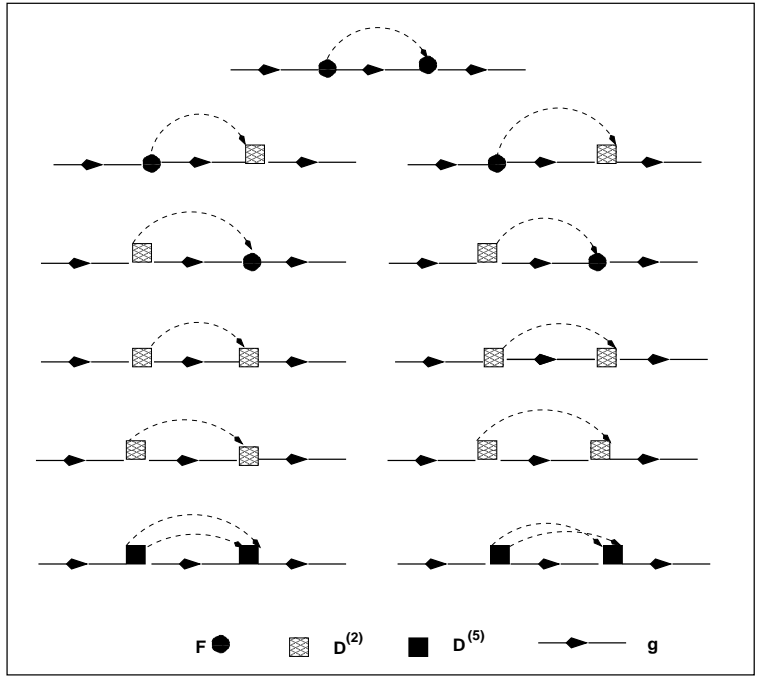}
\caption{The possible scattering diagrams for n=2.}
\label{fig2}
\end{figure}

A connected diagram to order `n' is then built up by stringing together (n+1) propagators
 associated with {\bf g}, connected by n-vertices with all fluctuation lines combined in pairs. For n=2 there are eleven possible scattering diagrams, whose algebraic form is given in Equation. (\ref{eq6}). These diagrams are shown in Fig. \ref{fig2}.
\begin{figure}
\centering
\includegraphics[height=20cm,width=13cm]{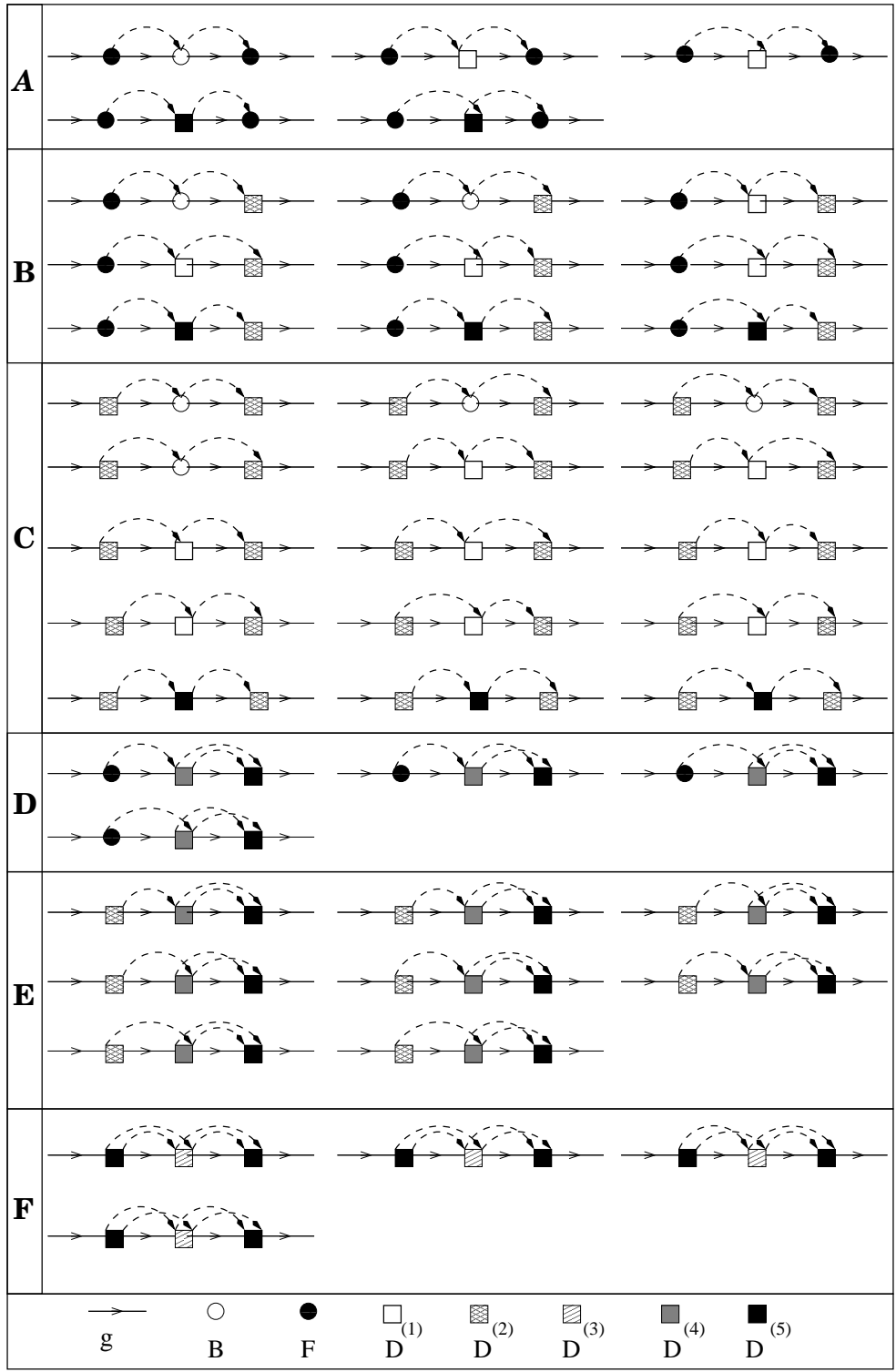}
\caption{The possible scattering diagrams for n=3.}
\label{fig3}
\end{figure}
For n=3, the possible kinds of scattering diagrams are shown in Fig. \ref{fig3}. Note that it involve terms with contribution from $\B, \D^{(1)}, \D^{(3)}$ and $\D^{(4)}$ as well. These scattering vertices can not sit either in the leftmost or in the rightmost positions, because then one of the associated pseudo-fermion green function line vanishes. In Fig. \ref{fig3}(A-F), each of the sections represents different distinct classes of diagrams. We call each sections as the topologically distinct scattering diagrams. There exists three more sections in addition to these (i.e.\ A-F) which are obtained simply by applying the reflection operators vertically to the diagrams in sections B, D and E. Hence in total, there exists 9 topologically distinct classes of scattering diagrams for n=3.

\begin{figure}
\centering
\includegraphics[height=20cm,width=13cm]{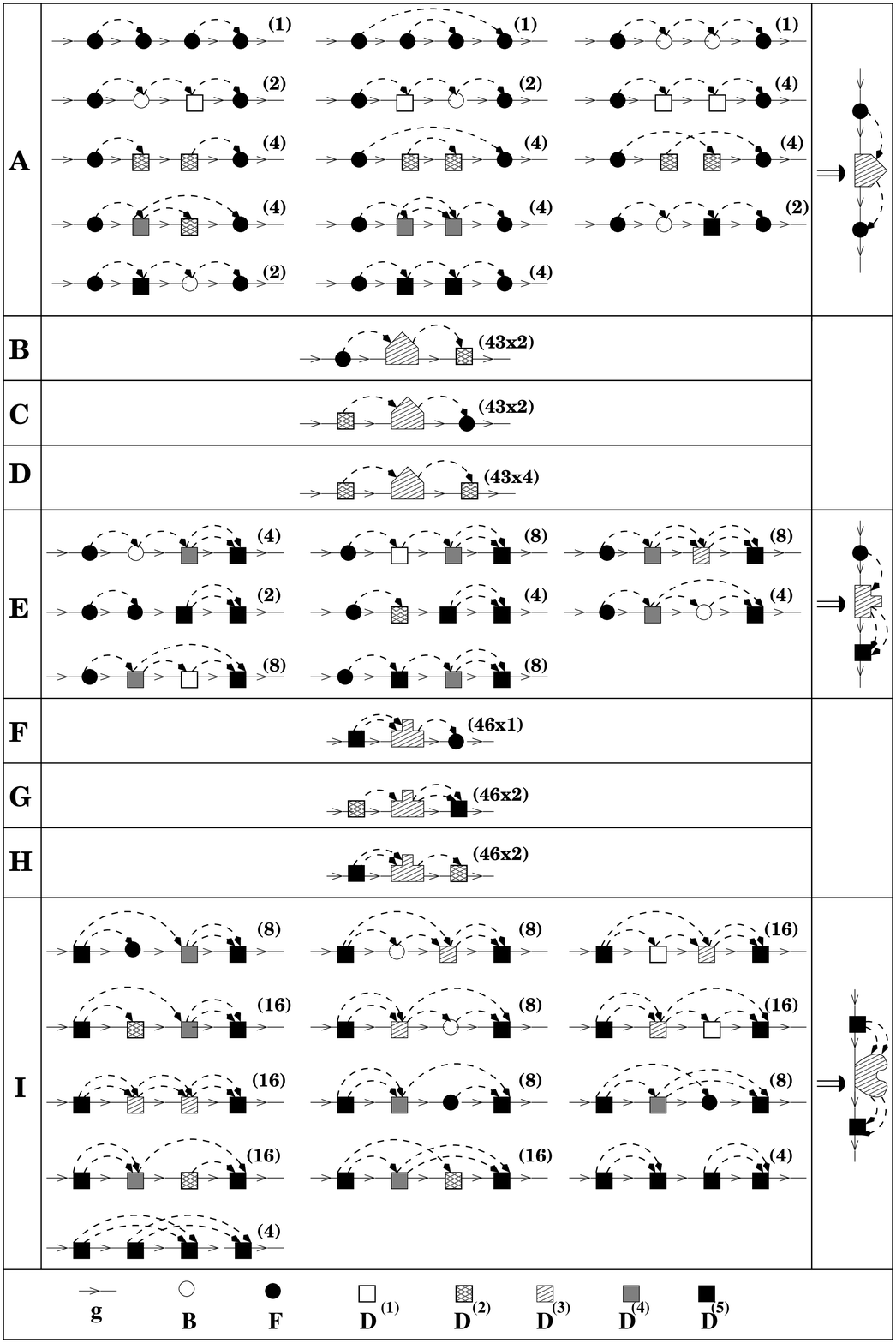}
\caption{The possible scattering diagrams for n=4.}
\label{fig4}
\end{figure}
For n=4, there are various classes of diagrams as shown in Fig. \ref{fig4}. In this figure we have not shown the scattering diagrams with all sorts of combinations of fluctuation lines, but label the multiplicity within the brackets () for each diagram which actually gives the number of possible ways of combining the fluctuation lines for that particular diagram. For the diagram in the last column of Fig. \ref{fig4}(A), the middle decoration indicates a collection of all those diagrams lying in between two vertices such that a single pseudo-fermion line goes out (\ from the left vertex\ ) and a single pseudo-fermion line comes in (\ at the right vertex\ ) for n=4. Similarly for the diagrams in the last  column of Fig. \ref{fig4}(E), the middle decoration indicates a collection of all those diagrams lying in between two vertices such that either (i) one pseudo-fermion line goes out (left vertex) and two pseudo-fermion line comes in (right vertex) or (ii) two pseudo-fermion line goes out (left vertex) and one pseudo-fermion line comes in (right vertex). While for the diagrams in the last column of Fig. \ref{fig4}(I), the middle decoration stands for a collection of all those diagrams lying in between two vertices such that two pseudo-fermion line goes out (from the left vertex) and two pseudo-fermion line comes in (at the right vertex). Each of these diagrams in the last column are said to be the topologically distinct classes of diagrams. Hence, overall there exists nine topologically distinct classes of diagrams for n=4 also.

Amongst the above sets of diagrams (\ for n=3 and n=4\ ), there exists three basic subsets namely
\begin{itemize}
\item A set of separable diagrams, which are those that can be broken into two along a propagator line without also breaking a pseudo-fermion line.
\item A set of non-separable, non-skeleton diagram which can not be broken into two, but the VCA green function in this diagram can be renormalized.
\item Skeleton diagrams involving all the crossed and complicated structured diagrams.
\end{itemize}
\begin{figure}[b]
\centering
\includegraphics[height=9.0cm,width=10cm]{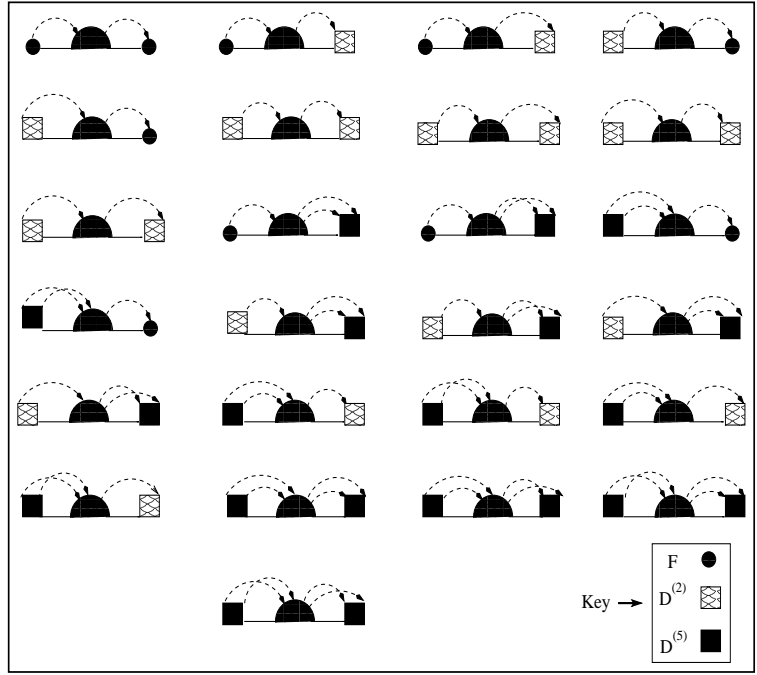}
\caption{Topological structure of the skeleton diagrams for the self-energy $\Sigma$. The central dark semicircle represents all possible arrangements of scattering vertices to all orders.}
\label{fig5}
\end{figure}

If we club together the contribution of {\sl all} the skeleton diagrams calling  this the
self-energy,  and allow {\sl all} phonon Green functions except
the left-most to be renormalized by the separable and non-separable, non-skeleton diagrams, we get the Dyson equation~:
\[ \ll {\mbf G}\gg \eq \g +    \g\ {\mathbf \Sigma} \ll {\mbf G}\gg \]
For homogeneous disorder we have shown earlier that we have translational symmetry in the full augmented
space \cite{gdma}. We can then take Fourier transform of the above equation to get~:

\[  
\ll\mbf{G}(\q,E)\gg \eq \mbf{g}(\q,E) + \mbf{g}(\q,E)\ \mbf{\Sigma}(\q,E)\ \ll\mbf{G}(\q,E)\gg
\label{dys1}
\]

The diagrams for the self-energy are skeleton diagrams {\sl all} of which have the structure as shown in Fig. \ref{fig5}. Each of these diagram starts or ends with any one of either $\F$, $\D^{(2)}$ or  $\D^{(5)}$ vertex. The central dark semicircle stands for all possible arrangements of scattering vertices to all orders.
\section{Effect of disorder on two-particle Green functions : Lattice
thermal conductivity}
\subsection{Configuration averaging of lattice thermal conductivity}
The Kubo formula which relates the optical conductivity to a current-current correlation function
is well established. The Hamiltonian contains a term $\sum_i {\mbf j}_i\cdot{\mbf A}({\r},t)$ which drives the electrical current. For thermal conductivity we do not have a similar term in the Hamiltonian which drives a heat current. The derivation of a Kubo formula in this situation requires an additional statistical hypothesis which states that a system in steady state has a space dependent {\sl local} temperature $T({\r}) = [\kappa_B \beta({\r})]^{-1}$. 
The expression for the heat current has been discussed in great detail by Hardy \cite{hardy} and 
Allen and Feldman \cite{af}. The readers are refereed to these papers
for the details of calculation. The matrix element of the heat current in the basis of the eigenfunctions of the Hamiltonian is given by~:

\begin{equation}
{\mbf S}_{\gamma\gamma^\prime}^\mu (\k)\ =\ \frac{\hbar}{2}\ \left(\rule{0mm}{3mm} \omega_{\k\gamma}+\omega_{\k\gamma^\prime}\right)\ {\mbf v}^\mu_{\gamma\gamma^\prime}(\k) ,
\end{equation}

\n where, the phonon group velocity ${\mbf v}_{\gamma\gamma^\prime}(\k)$ is given by

\begin{eqnarray}
{\mbf v}_{\gamma\gamma^\prime} & = & \frac{i}{2\sqrt{\omega_{\k\gamma}\omega_{\k\gamma^\prime}}}\ 
\sum_\mu\sum_\nu \epsilon^\mu_\gamma(\k)\ \epsilon^\nu_{\gamma^\prime}(\k)\nonumber \left( \sum_{\mbf R_{ij}}\ \frac{\Phi^{\mu\nu}({\mbf R_{ij}})}
{\sqrt{M_iM_j}}\ \right){\mbf R}_{ij}\  e^{i\k\cdot{\mbf R}_{ij}},\nonumber\\ 
 & = & \frac{1}{2\sqrt{\omega_{\k\gamma}\omega_{\k\gamma^\prime}}}\ \sum_\mu\sum_\nu \epsilon^\mu_\gamma(\k)\ {\mbf \nabla}_\k D^{\mu\nu}(\k) \ \epsilon^\nu_{\gamma^\prime}(\k)\nonumber ,\\
\end{eqnarray}

\n here i,j label the atoms,$\gamma,\gamma^\prime$ label the various modes of vibration, $\mu$, $\nu$ label the Cartesian directions, $\omega_{\k\gamma},\omega_{\k\gamma^\prime}$ are the eigen frequencies,  $\epsilon^\mu_\gamma(\k), \epsilon^\nu_{\gamma^\prime}(\k)$ are the polarization vectors associated with the $\gamma$-th and $\gamma^{\prime}$-th mode of vibration.\ $D^{\mu\nu}(\k)$ is the Fourier transform of mass scaled dynamical matrix.

We shall consider the case where the temperature gradient is uniform within the system. 
The linear heat current response is related to the temperature gradient  field 
via a generalized susceptibility :

\[ \langle S^{\mu}(t) \rangle = - \sum_{\nu}\int_{-\infty}^{\infty} \ dt'\ \kappa^{\mu\nu}(t-t')\  
{\mbf \nabla}^\nu \delta T(t),  \]
\noindent The generalized susceptibility is the thermal conductivity in this case and 

\[ 
\kappa^{\mu\nu}(\tau) =  \Theta(\tau)\ \frac{1}{T}\int_0^\beta\  d\lambda \langle  S^{\mu}(-i\hbar\lambda),S^{\nu}(\tau) \rangle ,
\]

$\Theta(\tau)$ is the Heaviside step function,
and
\[
S(-i\hbar\lambda)\ =\ e^{\lambda H}\ S\ e^{-\lambda H}.
\]

\n $\langle\ \rangle$ on the right-hand side stands for thermal averaging over states in the absence of the temperature gradient. 
The above equation can be rewritten in the form of a Kubo-Greenwood expression 
\[
\kappa^{\mu\nu}(\omega, T)  = \ \kappa^{\mu\nu}_I(\omega, T) \ +\ \kappa^{\mu\nu}_{II}(\omega, T) \]

\begin{eqnarray}
\fl \kappa^{\mu\nu}_I (\omega, T) = \frac{\pi}{T}\ \int \frac{d^3\k}{8\pi^3}\ \sum_{\gamma,\gamma^\prime\ne\gamma}\ \frac{\langle n_{\k\gamma^\prime}\rangle-\langle n_{\k\gamma}\rangle}{\hbar(\omega_{\k\gamma}-\omega_{\k\gamma^\prime})} 
{\mbf S}^\mu_{\gamma\gamma^\prime}(\k){\mbf S}^\nu_{\gamma^\prime\gamma}(\k) \ \delta(\omega_{\k\gamma}-\omega_{\k\gamma^\prime}-\omega) 
\label{eqint}
\end{eqnarray}
\begin{eqnarray}
\fl \kappa^{\mu\nu}_{II}(\omega, T) = \frac{1}{\kappa_B T^2} \left[ \rule{0mm}{4mm}\left\{ \int\frac{d^3\k}{8\pi^3} \sum_\gamma \langle n_{\k\gamma}\rangle\ {\mbf S}^\mu_{\gamma\gamma}(\k)\right\}
 \left\{ \int\frac{d^3\k}{8\pi^3}\sum_\gamma\langle n_{\k\gamma}\rangle\ {\mbf S}^\nu_{\gamma\gamma}(\k)\right\}\right.\nonumber\\
  \left. - \kappa_B T \int\frac{d^3\k}{8\pi^3}
 \sum_\gamma \frac{\partial\langle n_{\k\gamma}\rangle }
{\partial(\hbar\omega_{\k\gamma})}\ S^\mu_{\gamma\gamma}(\k)\ S^\nu_{\gamma\gamma}(\k) \right]\delta(\omega) 
\end{eqnarray}
where $\langle n_{\k\gamma}\rangle = (e^{\beta\hbar\omega_{\k\gamma}}-1)^{-1}$ is the equilibrium Bose Einstein distribution function and T is the absolute temperature.
For an isotropic response, equation. (\ref{eqint}) for inter-band transition  can be expressed as
\[
 \fl\kappa_I(\omega, T)= \frac{\pi}{3T}\sum_{\mu} \int d\omega^\prime \int\frac{d^3\k}{8\pi^3} \sum_{\gamma,\gamma^{\prime}}
\widehat{\mbf S}^\mu_{\gamma\gamma^\prime}(\k, T) 
 \widehat{\mbf S}^\mu_{\gamma^{\prime} \gamma}(\k, T) \delta(\omega^\prime - \omega_{\k\gamma^\prime})\delta(\omega^\prime+\omega-\omega_{\k\gamma}) 
\]

\n where 

\[
\widehat{\mbf S}^\mu_{\gamma\gamma^\prime}(\k,T)\ =\ \sqrt{\left|\frac{\langle n_{\k\gamma^\prime}\rangle-\langle n_{\k\gamma}\rangle}{\hbar(\omega_{\k\gamma}-\omega_{\k\gamma^\prime})}\right|}\ {\mbf S}^\mu_{\gamma\gamma^\prime}(\k).
\]

We may rewrite the above equation as 

\begin{eqnarray}
\fl\kappa_I(\omega,T) = \frac{1}{3\pi T}\ \sum_{\mu}\ \int d\omega^\prime\int\frac{d^3\k}{8\pi^3} \mbox{Tr} \left[\rule{0mm}{4mm}\ \widehat{\mbf S}^{\mu}(\k,T)
 \Im m \{ {\mbf G}(\k,\omega^\prime)\}\ \widehat{\mbf S}^{\mu}(\k,T)\ \Im m\{ {\mbf G}(\k,\omega^\prime+\omega)\} \right]\nonumber\\
\label{eq11}
\end{eqnarray}

The operator {\bf G}($\omega$) is the phonon Green operator $(M\omega^2{\mbf I}-{\mbf\Phi})^{-1}$. The Trace is invariant in different representations. For crystalline systems, usually the Bloch basis 
$\{\vert {\mbf k},\gamma\rangle\}$ is used. For disordered systems, prior to configuration averaging,
  it is more convenient to use the basis $\{\vert \k,\alpha\rangle\}$, where $\k$ is the reciprocal vector and $\alpha$
represents the coordinate axes directions.  We can transform from the mode basis to the coordinate basis by
using the transformation matrices $\Upsilon_{\gamma\alpha}(\k) =\ \epsilon^\alpha_\gamma(\k)$. For example 

\[ \widehat{\mbf S}_{\alpha\beta}^\mu(\k,T) \ =\     \Upsilon^{-T}_{\alpha\gamma}(\k)\ \widehat{\mbf S}^\mu_{\gamma\gamma^\prime}(\k,T)\ \Upsilon^{-1}_{\gamma^{\prime}\beta}(\k).
\]

\noindent If we define

\begin{equation} 
{\mbf \kappa}(z_1,z_2)  = \int\frac{d^3\k}{8\pi^3}\ \mbox{Tr} \left[\ \rule{0mm}{4mm}\widehat{\mbf S}\  {\mbf G}(\k,z_1)\ \widehat{\mbf S}\ {\mbf G}(\k,z_2)\right] .
\label{eq12}
\end{equation}

\noindent  then Equation. (\ref{eq11}) becomes,
\begin{eqnarray}
 \kappa_I(\omega,T) &=&  \frac{1}{12\pi T}\ \sum_{\mu}  \int d\omega^\prime \ \left[\rule{0mm}{4mm} {\mbf \kappa}^{\mu\mu}(\omega^{\prime -},\omega^{\prime +}+\omega) + {\mbf \kappa}^{\mu\mu}(\omega^{\prime +},\omega^{\prime -}+\omega)-\right.\nonumber\\ 
&&\left.   {\mbf \kappa}^{\mu\mu}(\omega^{\prime +},\omega^{\prime  +}+\omega) - {\mbf \kappa}^{\mu\mu}(\omega^{\prime -},\omega^{\prime  -}+\omega)\rule{0mm}{4mm}\right],
\label{eq13}
\end{eqnarray}

\noindent where

\[ 
f(\omega^\pm) = \lim_{\delta\rightarrow 0} f(\omega\pm i\delta).
\]

\noindent We have also used the herglotz analytic property of the Green operator   

\[
{\mbf G}(\omega+i\delta) = \Re e\left[\rule{0mm}{3mm}{\mbf G}(\omega)\right] \mns  i\ \mbox{sgn}(\delta)\ \Im m\left[\rule{0mm}{3mm}{\mbf G}(\omega)\right]. 
\]

For disordered materials, we shall be interested in obtaining the configuration averaged  response functions. This will require
the configuration averaging of quantities like $\kappa(z_1,z_2)$.
 For disordered materials, Equation. (\ref{eq13}) should be expressed as

\begin{eqnarray}
 \ll\kappa_I(\omega,T)\gg &=&  \frac{1}{12\pi T}\ \sum_{\mu}  \int d\omega^\prime \ \left\langle\left\langle\left[\rule{0mm}{4mm} {\mbf \kappa}^{\mu\mu}(\omega^{\prime -},\omega^{\prime +}+\omega) + {\mbf \kappa}^{\mu\mu}(\omega^{\prime +},\omega^{\prime -}+\omega)\right.\right.\right.\nonumber \\ &&\left.\left.\left.    - {\mbf \kappa}^{\mu\mu}(\omega^{\prime +},\omega^{\prime  +}+\omega)
 - {\mbf \kappa}^{\mu\mu}(\omega^{\prime -},\omega^{\prime  -}+\omega)\right]\rule{0mm}{4mm}\right\rangle\right\rangle,
\label{eq14}
\end{eqnarray}

Let us discuss the configuration averaging of the two particle green function of the kind $\kappa(z_1,z_2)$. The augmented space theorem immediately implies that

\begin{equation}
\ll {\mbf \kappa}(z_1,z_2)\gg = \mbox{Tr}\left\langle  \{\emptyset\} \left |\ \left[ \rule{0mm}{4mm} \right.\wt{\mbf S} \wt{\mbf G}(z_1)\ \wt{\mbf S}^{\dagger} \wt{\mbf G}(z_2)\right]\ \left.\rule{0mm}{4mm} \right | \{\emptyset\}\right\rangle
\label{eq15}
\end{equation}

The first thing to note about Equation. (\ref{eq15}) is that the right hand side is an average of four random functions whose fluctuations are correlated. The average of the product then involves the product of the averages and other contributions which come from averages taken in pairs, triplets and all four random functions.

Following a similar procedure as for a single particle green functions, the operator $\wt\S$ in the augmented space has the form

\begin{eqnarray}
\fl \wt{\S} = \sum_{R\alpha}\sum_{R'\alpha'}\left[\rule{0mm}{4mm} \ll \widehat{\S}\gg_{R\alpha,R'\alpha'}\wt{I}+{\S}^{(1)}_{R\alpha,R'\alpha'}\left({ p}^\downarrow_R + { p}^\downarrow_{R'}\right) + 
 {\S}^{(2)}_{R\alpha,R'\alpha'} \left({\cal T}^{\uparrow\downarrow}_R + {\cal T}^{\uparrow\downarrow}_{R'}\right)+\right.\nonumber\\ 
\fl\phantom{XXXX} \left. {\S}^{(3)}_{R\alpha,R'\alpha'} p^\downarrow_R\otimes p^\downarrow_{R'} + 
 {\S}^{(4)}_{R\alpha,R'\alpha'}
 \left(p^\downarrow_R\otimes {\cal T}^{\uparrow\downarrow}_{R'}+p^\downarrow_{R'} \otimes{\cal T}^{\uparrow\downarrow}_{R}\right)+{\S}^{(5)}_{R\alpha,R'\alpha'} {\cal T}^{\uparrow\downarrow}_R\otimes{\cal T}^{\uparrow\downarrow}_{R'}\right]\otimes T_{RR'},\nonumber\\ 
\label{eq16}
\end{eqnarray}
\n where

\begin{eqnarray*}
\S^{(1)}&=&(y-x)\ \widehat{\S}^{(1)},\ \  \S^{(2)} = \sqrt{xy}\ \widehat{\S}^{(1)}, \\ \S^{(3)} &=& (y-x)^2\ \widehat{\S}^{(2)},\ \ \S^{(4)} =\sqrt{xy}\ (y-x)\ \widehat{\S}^{(2)},\\ \S^{(5)} &=& xy\ \widehat{\S}^{(2)}.
\end{eqnarray*}

\n and

\begin{eqnarray*}
 \widehat{\S}^{(1)}_{R\alpha,R'\alpha^{\prime}} &=& x \left( \widehat{\S}^{AA}_{R\alpha,R'\alpha^{\prime}}- \widehat{\S}^{AB}_{R\alpha,R'\alpha^{\prime}}\right) -  y \left( \widehat{\S}^{BB}_{R\alpha,R'\alpha^{\prime}}-\widehat{\S}^{BA}_{R\alpha,R'\alpha^{\prime}}\right)\\
 \widehat{\S}^{(2)}_{R\alpha,R'\alpha^{\prime}}&=&\widehat{\S}^{AA}_{R\alpha,R'\alpha^{\prime}} + \widehat{\S}^{BB}_{R\alpha,R'\alpha^{\prime}} - \widehat{\S}^{AB}_{R\alpha,R'\alpha^{\prime}} - \widehat{\S}^{BA}_{R\alpha,R'\alpha^{\prime}}. 
\end{eqnarray*}

\begin{figure}[t]
\centering
\includegraphics[height=9.0cm,width=10cm]{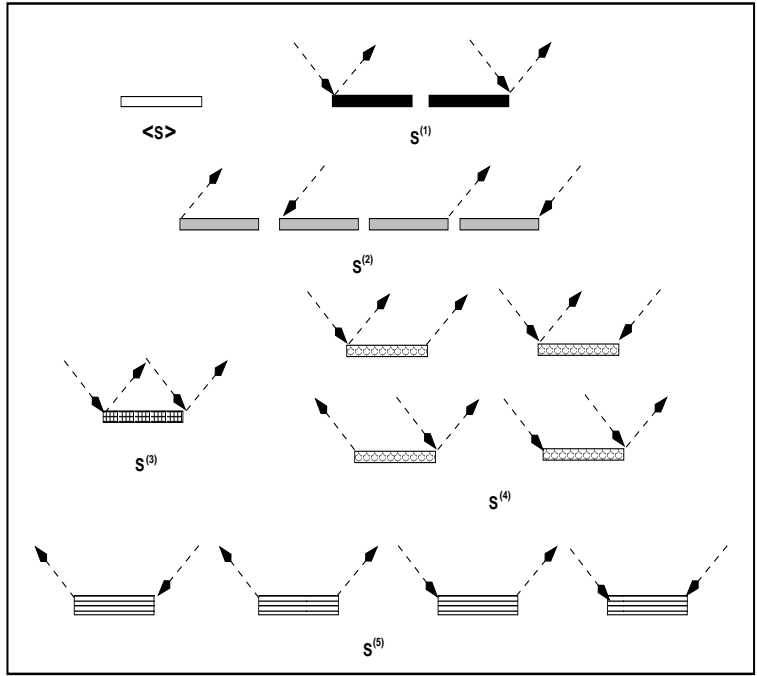}
\caption{The scattering vertices associated with the random current terms.}
\label{fig6}
\end{figure}
\subsection{The disorder renormalized current}

\begin{figure}[t]
\centering
\includegraphics[height=4.0cm,width=7cm]{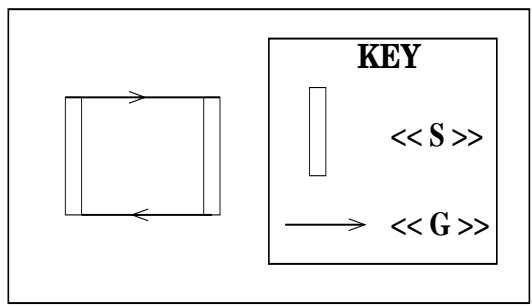}
\caption{The VCA or zero-th order approximation for $\ll\kappa(z_1,z_2)\gg$.}
\label{fig7}
\end{figure}

We now start to set up the scattering diagrams for the thermal conductivity.
A look at Equation. (\ref{eq16}) shows us that the first term is the averaged VCA current. This term is absorbed in the unscattered part of the phonon green function and leads to the zero-th order approximation. Equation. (\ref{eq16}) looks very similar to Equation. (\ref{eq4}) from the operator point of view. The only difference is that the former equation arises due to the disorder in heat currents while the latter equation due to the disorder in the dynamical matrix. 
Exactly as before we can associate scattering vertices with the terms in $\wt{\S}$. Fig. \ref{fig6} shows sixteen different scattering vertices arising out of the equation (\ref{eq16}). 
Let us now discuss how the scattering diagrams are set up and then examine
them.  The rule for obtaining the diagrams for the correlation function $\ll\kappa(z_1,z_2)\gg$  is as follows : Take any two current diagrams from Fig. \ref{fig6} and two propagators and join them end to end. Now join the configuration fluctuation lines (shown as dashed arrows) in all possible ways. The zero-th order approximation for $\ll\kappa(z_1,z_2)\gg$ can be shown diagrammatically as in Fig. \ref{fig7}. The most dominant contribution comes from this particular diagram. Here the two current terms are the averaged current, and all configuration fluctuation decorations renormalize only the two phonon propagators. The bold propagators in this diagram are fully scattering renormalized propagators corresponding to the configuration averaged green function. The contribution of this term to the correlation function $\ll\kappa(z_1,z_2)\gg$  is

\begin{equation} 
 \int\frac{d^3\k}{8\pi^3}\ \ll{\mbf {\widehat S}(\k)}\gg  \ll{\mbf G}(\k,z_1)\gg \ll{\mbf {\widehat S}(\k)}\gg \ll{\mbf G}(\k,z_2)\gg.
\label{eq17}
\end{equation}

\begin{figure}[b]
\centering
\includegraphics[height=9.0cm,width=11cm]{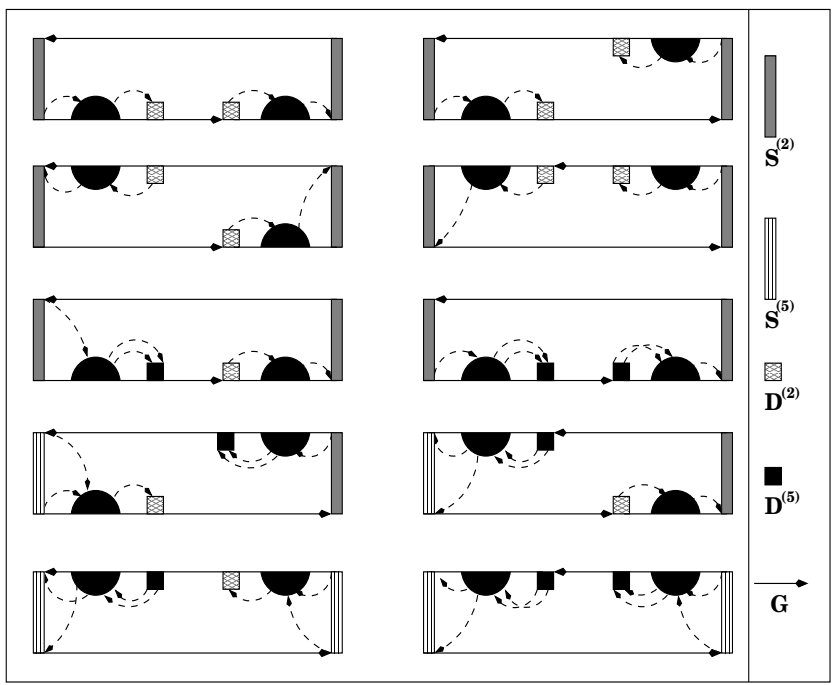}
\caption{Few examples of scattering diagrams where no disorder line joins the two phonon propagators.}
\label{fig8}
\end{figure}
The rest of the terms in Equation. (\ref{eq16}) give rise to scattering. We shall now focus on the main correction terms to the expression in Equation. (\ref{eq17}). These are the correction terms to the averaged current which, as we will show, are closely related to the self-energies. The first type of scattering diagrams are those in which no disorder propagator (shown as the dashed lines) joins either two phonon propagators or two of the current lines directly. Fig. (\ref{fig8}) shows few such scattering diagrams. These sets of diagrams may be clubbed together and renormalized in a form which will consist of two fully renormalized phonon propagators connected at the two ends by a new form of the renormalized current. This new form of the renormalized current may be obtained in the following way.

\begin{figure}[b]
\centering
\includegraphics[height=16.0cm,width=14cm]{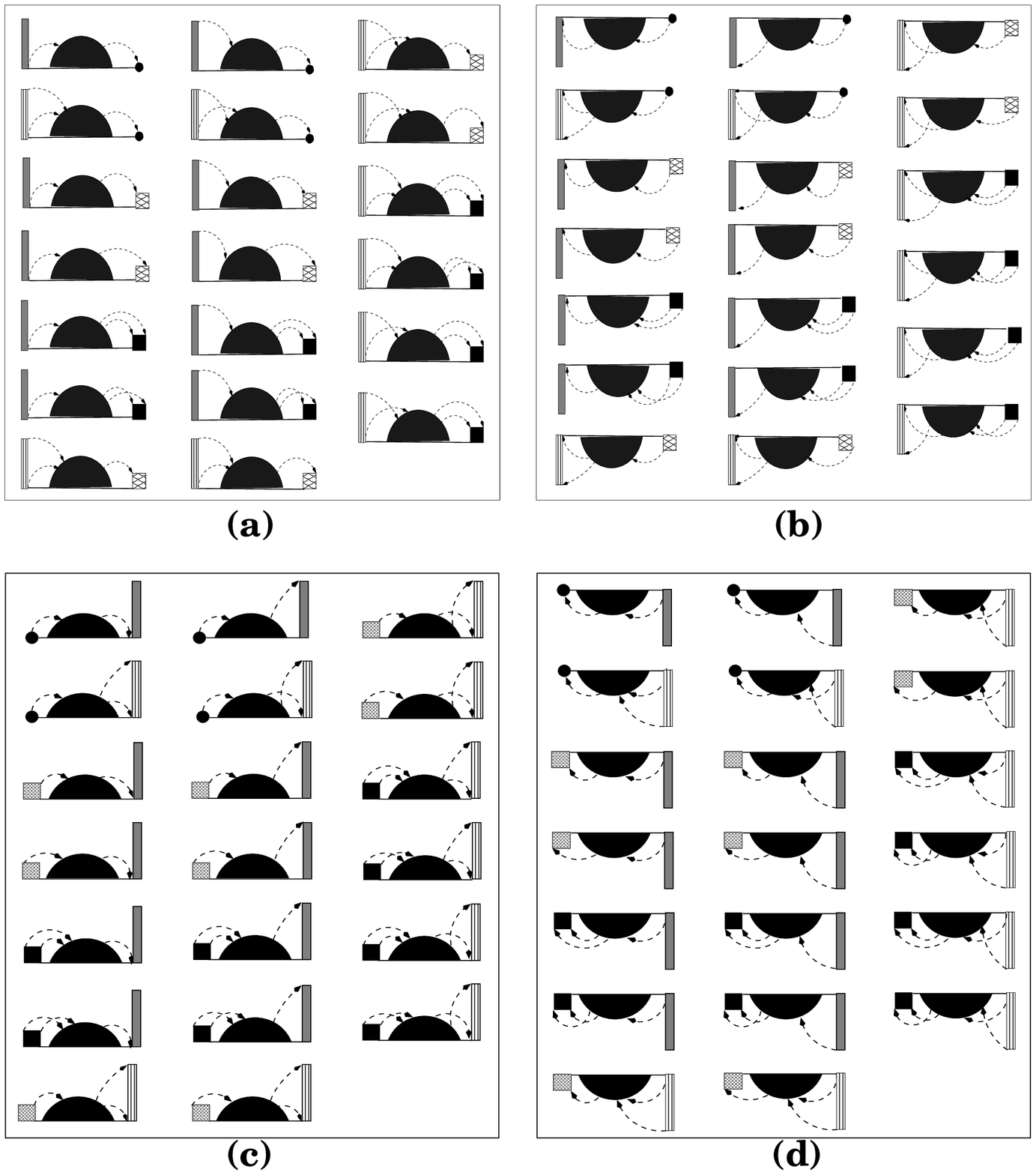}
\caption{Scattering diagrams contributing to effective heat current.}
\label{fig9}
\end{figure}

Fig. \ref{fig8} clearly shows that these types of diagrams are made out of a left $\it renormalized$ current diagram chosen out of any one of the diagrams from (a,b) in Fig. \ref{fig9} and one right $\it renormalized$ current diagram from any one of the diagram (c,d) of Fig. \ref{fig9} connected by two $\it renormalized$ propagators.

Let us now obtain expressions for the renormalized currents. A careful look at the self-energy diagrams (See Fig. \ref{fig5}) shows that all self-energy diagrams have the structure
\begin{equation}
{\mathbf\Sigma(\k,z)}=\Delta(\k,z)\ \  {\mathbf\Omega(\k,z)}\ \ \Delta(\k,z),
\label{eq18}
\end{equation}
Where ${\mathbf\Omega(\k,z)}$ is the Fourier transform of
\begin{equation}
{\Omega_{RR^{\prime}}(z)}=\sum_{R_{1}R_{2}}G_{RR_{1}}(z)\ P_{R_{1}R_{2}}^{RR^{\prime}}(z)\ G_{R_{2}R^{\prime}}(z),
\end{equation}
\n and
\[
\Delta(\k,z)=\F\ z^{2}+2\ \D^{(2)}(\k)+2\ \D^{(5)}(\k).
\]
 In the above equation, the quantity `${\it P}$' stands for the central dark semicircle of Fig. \ref{fig5} which represents all possible arrangements of scattering vertices to all orders.

\begin{figure}[h]
\centering
\includegraphics[height=9.0cm,width=9cm]{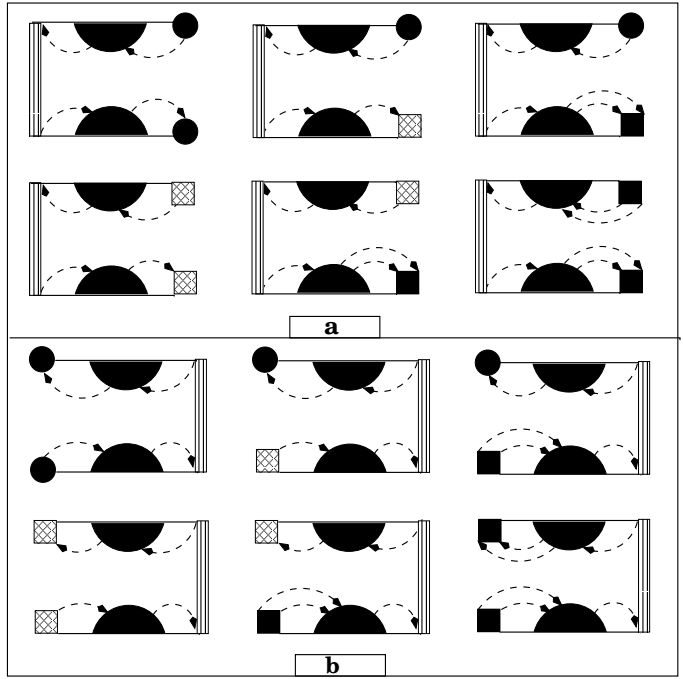}
\caption{The scattering diagrams associated with joint fluctuations of one current term and two propagators.}
\label{fig10}
\end{figure}

If we compare the diagrams of Fig. \ref{fig9}(a) with the diagrams for the self energy Fig. \ref{fig5}, we note that the only difference between the two is that the left most scattering vertex is replaced by a very similar current term. In the diagrams of Fig. \ref{fig9}(a), the left most diagonal terms similar to the vertex $\F$ of Fig. \ref{fig1} is off course missing. The contribution of such diagrams may be written in a mathematical form as 
\[
\left(2\ \S^{(2)}(\k)+2\ \S^{(5)}(\k)\right)\ \Omega(\k,z)\ \Delta(\k,z).
\] 
which may be expressed in terms of the self energy `$\mathbf\Sigma$' by using Equation. (\ref{eq18}) as
\[
\left(2\ \S^{(2)}(\k)+2\ \S^{(5)}(\k)\right)\ \left[\rule{0mm}{4mm}\Delta(\k,z_1)\right]^{-1}\ \mathbf\Sigma(\k,z_1).
\]
The contribution of the diagrams in Fig. {\ref {fig9}}(b) is
\[
\mathbf\Sigma(\k,z_2)\ \left[\rule{0mm}{4mm}\Delta(\k,z_2)\right]^{-1}\left(2\ \S^{(2)}(\k)+2\ \S^{(5)}(\k)\right).
\] 
Similarly the contribution of the diagrams in Fig. \ref{fig9}(c) and Fig. \ref{fig9}(d) are respectively given by
\[
\fl \mathbf\Sigma(\k,z_1)\ \left[\rule{0mm}{4mm}\Delta(\k,z_1)\right]^{-1}\left(2\ \S^{(2)}(\k)+2\ \S^{(5)}(\k)\right)\ \&\ 
\left(2\ \S^{(2)}(\k)+2\ \S^{(5)}(\k)\right)\ \left[\rule{0mm}{4mm}\Delta(\k,z_2)\right]^{-1}\ \mathbf\Sigma(\k,z_2)
\]
The next most dominant disorder  corrections come from a group of diagrams which describe joint fluctuation of one current and two propagators. A few such diagrams are shown in Fig. \ref{fig10}.

The contributions of these diagrams [Fig. \ref{fig10}(a,b)] can also be expressed in terms of the self energy as,
\begin{eqnarray}
&&(a)\ \ \mathbf\Sigma(\k,z_2)\ \left[\rule{0mm}{4mm}\Delta(\k,z_2)\right]^{-1}\ \S^{(5)}(\k)\ \left[\rule{0mm}{4mm}\Delta(\k,z_1)\right]^{-1}\ \mathbf\Sigma(\k,z_1),\nonumber\\
&&(b)\ \ \mathbf\Sigma(\k,z_1)\ \left[\rule{0mm}{4mm}\Delta(\k,z_1)\right]^{-1}\ \S^{(5)}(\k)\ \left[\rule{0mm}{4mm}\Delta(\k,z_2)\right]^{-1}\ \mathbf\Sigma(\k,z_2).\nonumber
\end{eqnarray}
If we now gather all the contributions from these diagrams [\ From Fig. \ref{fig9}(a,b,c,d) and Fig. \ref{fig10}(a,b)~], we may define a renormalized current term as follows :

\begin{equation}
{\S}_{\mathrm{eff}}(\k,z_1,z_2)=\ll \widehat{\S}(\k)\gg + \Delta{\S}_1(\k,z_1,z_2) + \Delta{\S}_2(\k,z_1,z_2),
\end{equation}

\n where

\begin{eqnarray}
\Delta {\S}_1(\k,z_1,z_2) &=& 2\ \left(\rule{0mm}{4mm} {\S}^{(2)}(\k) + {\S}^{(5)}(\k)\right)\left[\rule{0mm}{4mm}\Delta{(\k,z_1)}\right]^{-1} {\mbf \Sigma}(\k,z_1)+\nonumber\\
&&{\mbf \Sigma}(\k,z_2)\left[\rule{0mm}{4mm}\Delta{(\k,z_2)}\right]^{-1}2 \left(\rule{0mm}{4mm} {\S}^{(2)}(\k) + {\S}^{(5)}(\k)\right),\phantom{x}\nonumber\\
\Delta {\S}_2(\k,z_1,z_2) &=& {\mbf \Sigma}(\k,z_2)\left[\rule{0mm}{4mm}\Delta{(\k,z_2)}\right]^{-1}{\S}^{(5)}(\k) \left[\rule{0mm}{4mm}\Delta{(\k,z_1)}\right]^{-1}{\mbf  \Sigma}(\k,z_1).\nonumber
\end{eqnarray}

The contribution of these disorder-renormalized currents and propagators to the correlation function is

\begin{eqnarray}
\fl\ll\kappa_{(1)}(z_{1},z_{2})\gg = \int\frac{d^3\k}{8\pi^3}\ \mathrm{Tr}\left[{\S}_{\mathrm{eff}}(\k,z_{1},z_{2})\ll {\mbf{G}}({\bf k},z_{1})\gg
 {\S}_{\mathrm{eff}}^{\dagger}(\k,z_{1},z_{2})\ll {\mbf{G}}({\bf k},z_{2})\gg\right].
\label{eq21}
\end{eqnarray}

\begin{figure}[h]
\centering
\includegraphics[height=9.0cm,width=13cm]{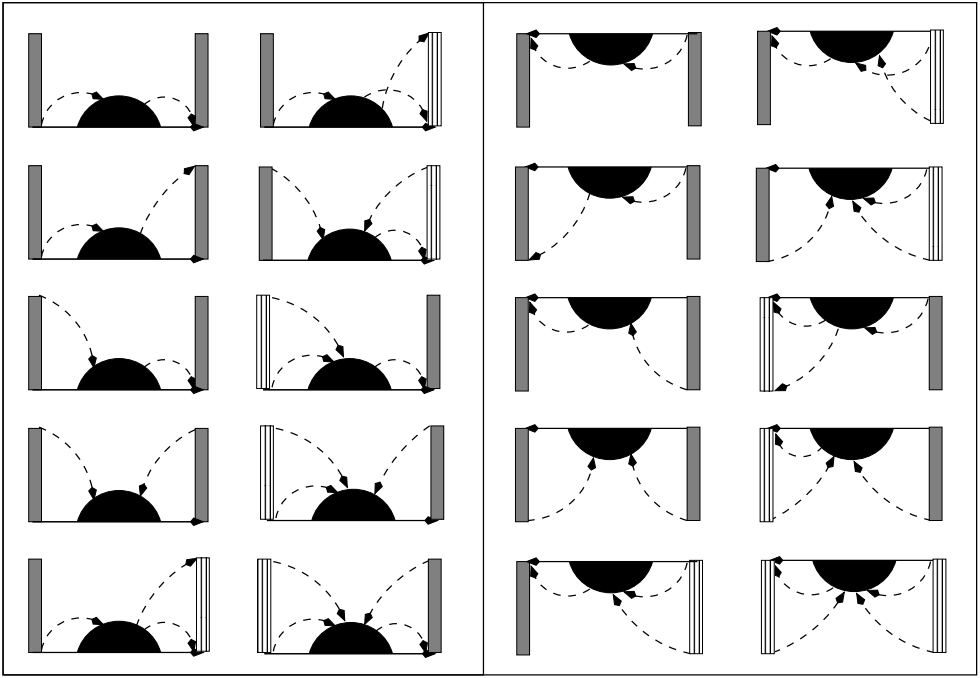}
\caption{The scattering diagrams associated with joint fluctuations of two current terms and one propagator.}
\label{fig11}
\end{figure}

We shall now discuss the disorder correction terms which involve joint fluctuations between the two current terms and one propagator. Few such diagrams are shown in Fig. \ref{fig11}. A close inspection of these diagrams shows that these are also related to the self energy diagrams with vertices at both ends replaced by currents. The corrections due to these terms can therefore be related to the self-energy as before.

The contribution of these diagrams to the correlation function is given by

\begin{eqnarray}
\fl \ll\kappa_{(2)}(z_{1},z_{2})\gg = 4 \int\frac{d^3\k}{8\pi^3}\ \mathrm{Tr}\left[\Delta{\S}_{3}(\k,z_{1})\ll {\mbf{G}}({\bf k},z_{2})\gg\ +\ \Delta{\S}_{4}(\k,z_{2})\ll {\mbf{G}}({\bf k},z_{1})\gg\right].
\label{eq22}
\end{eqnarray}

where

\begin{eqnarray*}
\fl\Delta {\S}_3(\k,z_1) = \left(\rule{0mm}{4mm} {\S}^{(2)}(\k) + {\S}^{(5)}(\k)\right)\left[\rule{0mm}{4mm}\Delta{(\k,z_1)}\right]^{-1} {\mbf \Sigma}(\k,z_1)
\left[\rule{0mm}{4mm}\Delta{(\k,z_1)}\right]^{-1} \left(\rule{0mm}{4mm} {\S}^{(2)}(\k) + {\S}^{(5)}(\k)\right)^{\dagger},\nonumber\\
\fl\Delta {\S}_4(\k,z_2) = \left(\rule{0mm}{4mm} {\S}^{(2)}(\k) + {\S}^{(5)}(\k)\right)\left[\rule{0mm}{4mm}\Delta{(\k,z_2)}\right]^{-1} {\mbf \Sigma}(\k,z_2)
\left[\rule{0mm}{4mm}\Delta{(\k,z_2)}\right]^{-1} \left(\rule{0mm}{4mm} {\S}^{(2)}(\k) + {\S}^{(5)}(\k)\right)^{\dagger}.
\end{eqnarray*}

In our earlier paper \cite{am3} on a similar problem, we have argued that these are the dominant disorder corrections to the average current. Intuitively we also expect the same to be true in the present case as well. It is important to note that that these corrections can be obtained from the self-energy and is therefore eminently computationally feasible in the case of realistic alloys, once we have a feasible method for obtaining the self-energy.

There are other scattering diagrams which are not related to the self-energy but rather to the vertex corrections. In these diagrams, a disorder line connects both the phonon propagators directly. We expect the corrections from these types of diagrams to be less dominant. For the sake of completeness, we shall indicate in detail how to obtain them within a ladder diagram approximation in the next section.

\subsection {The vertex correction}
\begin{figure}[t]
\centering
\includegraphics[height=9.0cm,width=11cm]{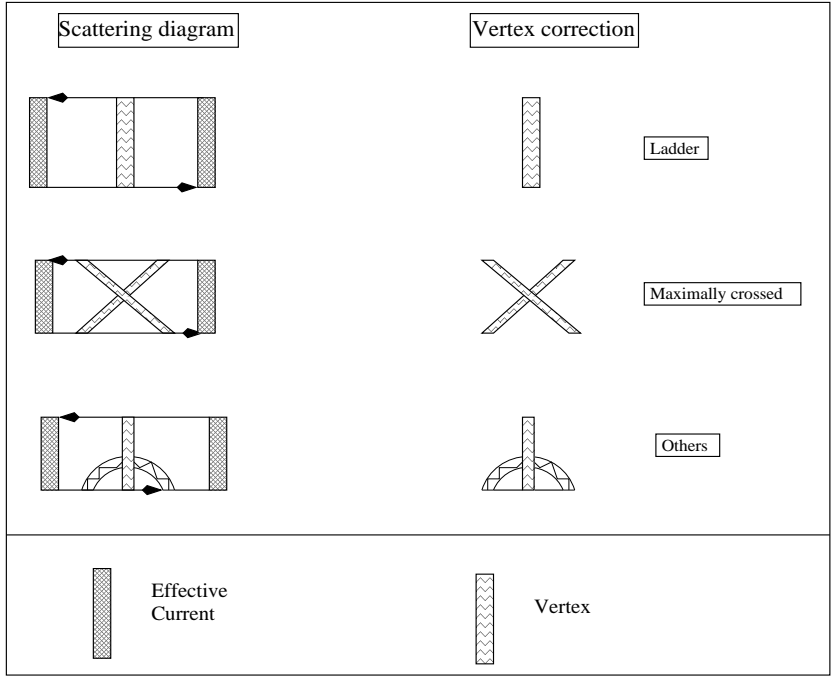}
\caption{The scattering diagrams leading to vertex correction.}
\label{fig12}
\end{figure}

\begin{figure}[t]
\centering
\includegraphics[height=19.0cm,width=13cm]{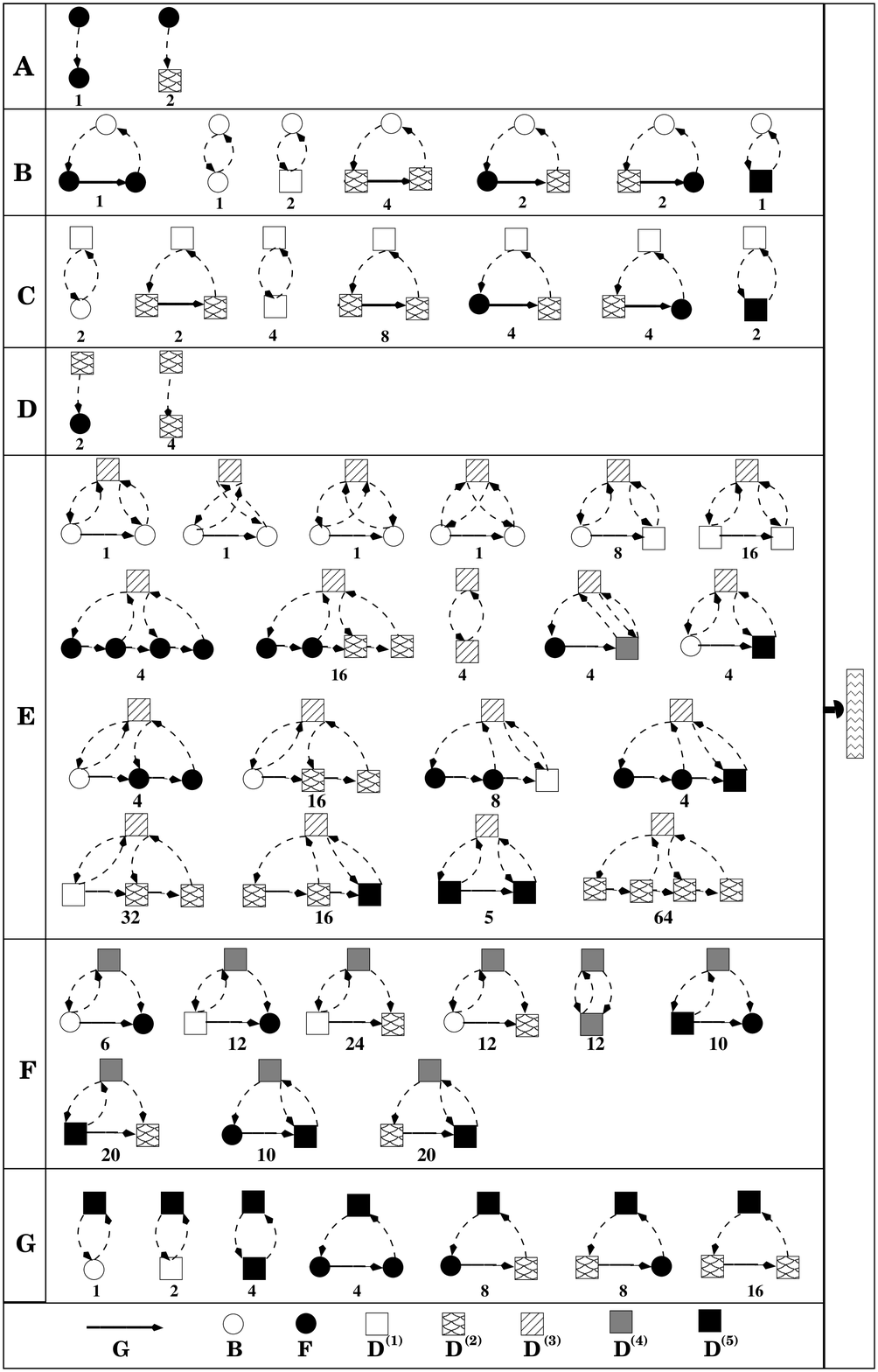}
\caption{The ladder scattering diagrams for the vertex correction.}
\label{fig13}
\end{figure}

The vertex corrections are basically those scattering diagrams in which disorder lines connect both the propagators directly. We have not yet incorporated these kinds of diagrams in the disorder renormalization. These types of diagrams arise due to the correlated propagation. The diagrams leading to the vertex corrections may be of different kinds {\it e.g.} ladder diagrams, maximally crossed diagrams etc. The ladder diagrams are those diagrams which are built out of repeated vertices shown on the first line of Fig. \ref{fig12}. These kinds of diagrams can be summed up to all orders. This is the disorder scattering version of the {\it random-phase approximation} (RPA) for the phonon-phonon scattering. The maximally crossed diagrams are those diagrams in which the ladder inserts between the crossed vertices. These types of diagrams are shown in the second line of Fig. \ref{fig12}.
Here we shall consider the ladder diagrams in detail and show how to obtain the contribution of these diagrams in terms of mathematical expression. We then sum these ladder diagrams to all orders.
The possible scattering diagrams for the ladder kind of vertex correction involving the vertices $\B, \F, \D^{(1)}$ to $\D^{(5)}$ are as shown in Fig. \ref{fig13}.

The contribution of seven categories (\ A-G\ ), shown in Fig. \ref{fig13}, of the ladder scattering diagrams in terms of mathematical expression are as given below.

{\sl\underline {Category A}}

\[
\hspace{-2.4cm} \left(W^{\gamma\delta}_{\alpha\beta}\right)_{A} = (z_1z_2)^2\ F_\alpha F_\gamma\ \delta_{\alpha\beta}\delta_{\gamma\delta} + 2\ z_2^2\ D^{(2)}_{\alpha\beta}\ F^\gamma\ \delta_{\gamma\delta}\nonumber
\]

{\sl\underline {Category B}}
\begin{eqnarray*}
\hspace{-2.4cm} \left(W^{\gamma\delta}_{\alpha\beta}\right)_{B} =({z_1}^{2}z_2)^{2} \left[\sum_{\nu^{\prime}\nu^{\prime\prime}}( F_{\alpha\nu^{\prime}} \delta_{\alpha\nu^{\prime}})\ G_{R\nu^{\prime},R\nu^{\prime\prime}}\ ( F_{\nu^{\prime\prime}\beta} \delta_{\nu^{\prime\prime}\beta})\right](B_{\nu\delta}\delta_{\nu\delta})\nonumber\\
\hspace{-1cm} + ({z_1}^{2}B_{\alpha\beta}\delta_{\alpha\beta}) ({z_2}^{2}B_{\nu\delta}\delta_{\nu\delta})+2\left[D^{(1)}_{\alpha\beta}({z_2}^{2}B_{\nu\delta}\delta_{\nu\delta} )\right]+4\left[\sum_{\nu^{\prime}\nu^{\prime\prime}} D^{(2)}_{\alpha\nu^{\prime}} G_{R\nu^{\prime},R\nu^{\prime\prime}}  D^{(2)}_{\nu^{\prime\prime}\beta} \right] \nonumber\\
\hspace{-1cm} \times({z_2}^{2}B_{\nu\delta}\delta_{\nu\delta}) + 2 \left[\sum_{\nu^{\prime}\nu^{\prime\prime}}({z_1}^{2} F_{\alpha\nu^{\prime}} \delta_{\alpha\nu^{\prime}}) G_{R\nu^{\prime},R\nu^{\prime\prime}}  D^{(2)}_{\nu^{\prime\prime}\beta} \right]({z_2}^{2}B_{\nu\delta}\delta_{\nu\delta}) \nonumber\\
\hspace{-1cm}2 \left[\sum_{\nu^{\prime}\nu^{\prime\prime}} D^{(2)}_{\alpha\nu^{\prime}}\ G_{R\nu^{\prime},R\nu^{\prime\prime}}\ ({z_1}^{2} F_{\nu^{\prime\prime}\beta} \delta_{\nu^{\prime\prime}\beta}) \right]({z_2}^{2}B_{\nu\delta}\delta_{\nu\delta}) +  D^{(5)}_{\alpha\beta}({z_2}^{2}B_{\nu\delta}\delta_{\nu\delta}).\nonumber\\
\end{eqnarray*}

{\sl\underline {Category C}}
\begin{eqnarray*}
\hspace{-2.4cm} \left(W^{\gamma\delta}_{\alpha\beta}\right)_{C} = 2\left[({z_1}^{2}B_{\alpha\beta}\delta_{\alpha\beta} ) D^{(1)}_{\nu\delta}\right] + 2 \left[\sum_{\nu^{\prime}\nu^{\prime\prime}}({z_1}^{2} F_{\alpha\nu^{\prime}} \delta_{\alpha\nu^{\prime}})\ G_{R\nu^{\prime},R\nu^{\prime\prime}}\ ({z_1}^{2} F_{\nu^{\prime\prime}\beta} \delta_{\nu^{\prime\prime}\beta})\right] D^{(1)}_{\nu\delta}\nonumber\\
\hspace{-1cm}\ +\ 4\ D^{(1)}_{\alpha\beta}\ D^{(1)}_{\nu\delta} + 8 \left[\sum_{\nu^{\prime}\nu^{\prime\prime}}\ D^{(2)}_{\alpha\nu^{\prime}} \ G_{R\nu^{\prime},R\nu^{\prime\prime}}\ D^{(2)}_{\nu^{\prime\prime}\beta} \right] D^{(1)}_{\nu\delta}\ +\ 4 \left[\sum_{\nu^{\prime}\nu^{\prime\prime}}({z_1}^{2} F_{\alpha\nu^{\prime}} \delta_{\alpha\nu^{\prime}})\right.\nonumber\\
\hspace{-0.7cm}\ \left. G_{R\nu^{\prime},R\nu{\prime\prime}}\  D^{(2)}_{\nu^{\prime\prime}\beta} \rule{0mm}{6mm}\right] D^{(1)}_{\nu\delta}\ + 2 \left[\sum_{\nu^{\prime}\nu^{\prime\prime}} D^{(2)}_{\alpha\nu^{\prime}}\ G_{R\nu^{\prime},R\nu^{\prime\prime}}\ ({z_1}^{2} F_{\nu^{\prime\prime}\beta} \delta_{\nu^{\prime\prime}\beta}) \right] D^{(1)}_{\nu\delta} 
\end{eqnarray*}

{\sl\underline {Category D}}

\begin{eqnarray*}
\hspace{-2.4cm} \left(W^{\gamma\delta}_{\alpha\beta}\right)_{D} = 2\left[({z_1}^{2}F_{\alpha\beta}\delta_{\alpha\beta} ) D^{(2)}_{\nu\delta}\right] + 4\ D^{(2)}_{\alpha\beta}\ D^{(2)}_{\nu\delta}
\end{eqnarray*}

{\sl\underline {Category E}}
\begin{eqnarray*}
\hspace{-2.4cm} \left(W^{\gamma\delta}_{\alpha\beta}\right)_{E} = 2 \left[\sum_{\nu^{\prime}\nu^{\prime\prime}}({z_1}^{2} B_{\alpha\nu^{\prime}} \delta_{\alpha\nu^{\prime}})\ G_{R\nu^{\prime},R\nu^{\prime\prime}}\ ({z_1}^{2} B_{\nu^{\prime\prime}\beta} \delta_{\nu^{\prime\prime}\beta})\right] D^{(3)}_{\nu\delta}+\ 8 \left[\sum_{\nu^{\prime}\nu^{\prime\prime}}({z_1}^{2} B_{\alpha\nu^{\prime}} \delta_{\alpha\nu^{\prime}})\right.\\
\hspace{-1.0cm}\ \left. G_{R\nu^{\prime},R\nu^{\prime\prime}}\  D^{(1)}_{\nu^{\prime\prime}\beta} \rule{0mm}{6mm}\right] D^{(3)}_{\nu\delta}\ + 16 \left[\sum_{\nu^{\prime}\nu^{\prime\prime}}\ D^{(1)}_{\alpha\nu^{\prime}} \ G_{R\nu^{\prime},R\nu^{\prime\prime}}\ D^{(1)}_{\nu^{\prime\prime}\beta} \right] D^{(3)}_{\nu\delta}+ \\
\hspace{-1cm}\ +\ 4 \left[\sum_{\nu_{1}\ldots\nu_{6}}({z_1}^{2} F_{\alpha\nu_{1}} \delta_{\alpha\nu_{1}})\ G_{R\nu_{1},R\nu_{2}}\ ({z_1}^{2} F_{\nu_{2}\nu_{3}} \delta_{\nu_{2}\nu_{3}})\ G_{R\nu_{3},R\nu_{4}}\ ({z_1}^{2} F_{\nu_{4}\nu_{5}} \delta_{\nu_{4}\nu_{5}})\right.\\
\hspace{-1.0cm}\ \left.  G_{R\nu_{5},R\nu_{6}}\ ({z_1}^{2} F_{\nu_{6}\beta} \delta_{\nu_{6}\beta})\ \rule{0mm}{6mm}\right] D^{(3)}_{\nu\delta}+ 16 \left[\sum_{\nu_{1}\ldots\nu_{6}}({z_1}^{2} F_{\alpha\nu_{1}} \delta_{\alpha\nu_{1}})\ G_{R\nu_{1},R\nu_{2}}\right. \\
\hspace{-1.0cm}\ \left. ({z_1}^{2} F_{\nu_{2}\nu_{3}} \delta_{\nu_{2}\nu_{3}})\ G_{R\nu_{3},R\nu_{4}}\ D^{(2)}_{\nu_{4}\nu_{5}}  G_{R\nu_{5},R\nu_{6}}\ D^{(2)}_{\nu_{6}\beta} \ \rule{0mm}{6mm}\right] D^{(3)}_{\nu\delta} + 4 D^{(3)}_{\alpha\beta}\ D^{(3)}_{\nu\delta}\\
\hspace{-1cm}\ + 4 \left[\sum_{\nu^{\prime}\nu^{\prime\prime}}({z_1}^{2} F_{\alpha\nu^{\prime}} \delta_{\alpha\nu^{\prime}})\ G_{R\nu^{\prime},R\nu^{\prime\prime}}\ D^{(4)}_{\nu^{\prime\prime}\beta} \right] D^{(3)}_{\nu\delta} + 4 \left[\sum_{\nu^{\prime}\nu^{\prime\prime}}({z_1}^{2} F_{\alpha\nu^{\prime}} \delta_{\alpha\nu^{\prime}}) \right.\\
\hspace{-1.0cm}\ \left. G_{R\nu^{\prime},R\nu^{\prime\prime}}\ D^{(5)}_{\nu^{\prime\prime}\beta} \rule{0mm}{6mm}\right] D^{(3)}_{\nu\delta} + 4 \left[\sum_{\nu_{1}\ldots\nu_{4}}({z_1}^{2} B_{\alpha\nu_{1}} \delta_{\alpha\nu_{1}})\ G_{R\nu_{1},R\nu_{2}}\ ({z_1}^{2} F_{\nu_{2}\nu_{3}} \delta_{\nu_{2}\nu_{3}})\right.\nonumber\\
\hspace{-1.0cm}\ \left. G_{R\nu_{3},R\nu_{4}}\ ({z_1}^{2} F_{\nu_{4}\delta} \delta_{\nu_{4}\delta})\rule{0mm}{6mm}\right] D^{(3)}_{\nu\delta} + 16 \left[\sum_{\nu_{1}\ldots\nu_{4}}({z_1}^{2} B_{\alpha\nu_{1}} \delta_{\alpha\nu_{1}})\ G_{R\nu_{1},R\nu_{2}}\ (D^{(2)}_{\nu_{2}\nu_{3}}) \right.\nonumber\\
\hspace{-1.0cm}\ \left. G_{R\nu_{3},R\nu_{4}}\ (D^{(2)}_{\nu_{4}\beta})\rule{0mm}{6mm}\right] D^{(3)}_{\nu\delta} + 8 \left[\sum_{\nu_{1}\ldots\nu_{4}}({z_1}^{2} F_{\alpha\nu_{1}} \delta_{\alpha\nu_{1}})\ G_{R\nu_{1},R\nu_{2}}\ ({z_1}^{2} F_{\nu_{2}\nu_{3}} \delta_{\nu_{2}\nu_{3}})\right.\nonumber\\
\hspace{-1.0cm}\ \left. G_{R\nu_{3},R\nu_{4}}\ (D^{(1)}_{\nu_{4}\beta})\rule{0mm}{6mm}\right] D^{(3)}_{\nu\delta} + 4 \left[\sum_{\nu_{1}\ldots\nu_{4}}({z_1}^{2} F_{\alpha\nu_{1}} \delta_{\alpha\nu_{1}})\ G_{R\nu_{1},R\nu_{2}}\ ({z_1}^{2} F_{\nu_{2}\nu_{3}} \delta_{\nu_{2}\nu_{3}})\right.\nonumber\\
\hspace{-1.0cm}\ \left. G_{R\nu_{3},R\nu_{4}}\ (D^{(5)}_{\nu_{4}\beta})\rule{0mm}{6mm}\right] D^{(3)}_{\nu\delta}  + 32 \left[\sum_{\nu_{1}\ldots\nu_{4}}(D^{(1)}_{\alpha\nu_{1}})\ G_{R\nu_{1},R\nu_{2}}\ (D^{(2)}_{\nu_{2}\nu_{3}})G_{R\nu_{3},R\nu_{4}}\ (D^{(2)}_{\nu_{4}\beta})\rule{0mm}{6mm}\right] \nonumber\\
\hspace{-1.0cm}\  D^{(3)}_{\nu\delta}  + 16 \left[\sum_{\nu_{1}\ldots\nu_{4}}(D^{(2)}_{\alpha\nu_{1}})\ G_{R\nu_{1},R\nu_{2}}\ (D^{(2)}_{\nu_{2}\nu_{3}})G_{R\nu_{3},R\nu_{4}}\ (D^{(5)}_{\nu_{4}\beta})\right]D^{(3)}_{\nu\delta}\nonumber\\
\hspace{-1.0cm}\ + 5 \left[\sum_{\nu^{\prime}\nu^{\prime\prime}}(D^{(5)}_{\alpha\nu^{\prime}})\ G_{R\nu^{\prime},R\nu^{\prime\prime}}\ (D^{(5)}_{\nu^{\prime\prime}\beta})\right]D^{(3)}_{\nu\delta}+ 64 \left[\sum_{\nu_{1}\ldots\nu_{6}}(D^{(2)}_{\alpha\nu_{1}})\ G_{R\nu_{1},R\nu_{2}}\ (D^{(2)}_{\nu_{2}\nu_{3}})\right.\nonumber\\
\hspace{-1.0cm}\ \left. G_{R\nu_{3},R\nu_{4}}\ (D^{(2)}_{\nu_{4}\nu_{5}}) G_{R\nu_{5},R\nu_{6}}\ (D^{(2)}_{\nu_{6}\beta})\rule{0mm}{6mm}\right]D^{(3)}_{\nu\delta}
\end{eqnarray*}

{\sl \underline {Category F}}
\begin{eqnarray*}
\hspace{-2.4cm} \left(W^{\gamma\delta}_{\alpha\beta}\right)_{F} = 6 \left[\sum_{\nu^{\prime}\nu^{\prime\prime}}({z_1}^{2} B_{\alpha\nu^{\prime}} \delta_{\alpha\nu^{\prime}})\ G_{R\nu^{\prime},R\nu^{\prime\prime}}\ ({z_1}^{2} F_{\nu^{\prime\prime}\beta} \delta_{\nu^{\prime\prime}\beta})\right] D^{(4)}_{\nu\delta} + 12 \left[\sum_{\nu^{\prime}\nu^{\prime\prime}}(D^{(1)}_{\alpha\nu^{\prime}})\ G_{R\nu^{\prime},R\nu^{\prime\prime}}\right.\\
\hspace{-1.0cm}\ \left. ({z_1}^{2} F_{\nu^{\prime\prime}\beta} \delta_{\nu^{\prime\prime}\beta})\rule{0mm}{6mm}\right] D^{(4)}_{\nu\delta} + 24 \left[\sum_{\nu^{\prime}\nu^{\prime\prime}}(D^{(1)}_{\alpha\nu^{\prime}})\ G_{R\nu^{\prime},R\nu^{\prime\prime}}\ (D^{(2)}_{\nu^{\prime\prime}\beta})\right] D^{(4)}_{\nu\delta}\\
\hspace{-1.0cm}\ + 12 \left[\sum_{\nu^{\prime}\nu^{\prime\prime}}({z_1}^{2} B_{\alpha\nu^{\prime}} \delta_{\alpha\nu^{\prime}})\ G_{R\nu^{\prime},R\nu^{\prime\prime}}(D^{(2)}_{\nu^{\prime\prime}\beta})\right] D^{(4)}_{\nu\delta} + 12\ D^{(4)}_{\alpha\beta}\ D^{(4)}_{\nu\delta}\\
\hspace{-1.0cm}\ + 10 \left[\sum_{\nu^{\prime}\nu^{\prime\prime}}(D^{(5)}_{\alpha\nu^{\prime}})\ G_{R\nu^{\prime},R\nu^{\prime\prime}}\ ({z_1}^{2} F_{\nu^{\prime\prime}\beta} \delta_{\nu^{\prime\prime}\beta})\right] D^{(4)}_{\nu\delta} + 20 \left[\sum_{\nu^{\prime}\nu^{\prime\prime}}(D^{(5)}_{\alpha\nu^{\prime}})\ G_{R\nu^{\prime},R\nu{\prime\prime}}\right.\\
\hspace{-1.0cm}\ \left. \times\ (D^{(2)}_{\nu^{\prime\prime}\beta})\right] D^{(4)}_{\nu\delta} + 10 \left[\sum_{\nu^{\prime}\nu^{\prime\prime}}({z_1}^{2} F_{\alpha\nu^{\prime}} \delta_{\alpha\nu^{\prime}})\ G_{R\nu^{\prime},R\nu^{\prime\prime}}(D^{(5)}_{\nu^{\prime\prime}\beta})\right] D^{(4)}_{\nu\delta}\\
\hspace{-1.0cm}\ +\ 20 \left[\sum_{\nu^{\prime}\nu^{\prime\prime}}(D^{(2)}_{\alpha\nu^{\prime}})\ G_{R\nu^{\prime},R\nu^{\prime\prime}} (D^{(5)}_{\nu^{\prime\prime}\beta})\right] D^{(4)}_{\nu\delta}
\end{eqnarray*}

{\sl\underline {Category G}}
\begin{eqnarray*}
\hspace{-2.4cm} \left(W^{\gamma\delta}_{\alpha\beta}\right)_{G}= ({z_1}^{2} B_{\alpha\beta} \delta_{\alpha\beta})\ D^{(5)}_{\nu\delta} + 2 D^{(1)}_{\alpha\beta} D^{(5)}_{\nu\delta} + 4 D^{(5)}_{\alpha\beta} D^{(5)}_{\nu\delta} + 4 \left[\sum_{\nu^{\prime}\nu^{\prime\prime}}({z_1}^{2} F_{\alpha\nu^{\prime}} \delta_{\alpha\nu^{\prime}})\ G_{R\nu^{\prime},R\nu^{\prime\prime}}\right.\\
\hspace{-1.0cm}\ \left. ({z_1}^{2} F_{\nu^{\prime\prime}\beta} \delta_{\nu^{\prime\prime}\beta})\rule{0mm}{6mm}\right] D^{(5)}_{\nu\delta} + 8 \left[\sum_{\nu^{\prime}\nu^{\prime\prime}}({z_1}^{2} F_{\alpha\nu^{\prime}} \delta_{\alpha\nu^{\prime}})\ G_{R\nu^{\prime},R\nu^{\prime\prime}} (D^{(2)}_{\nu^{\prime\prime}\beta})\right] D^{(5)}_{\nu\delta}\\
\hspace{-1.0cm}\ + 8 \left[\sum_{\nu^{\prime}\nu^{\prime\prime}} (D^{(2)}_{\alpha\nu^{\prime}})\ G_{R\nu^{\prime},R\nu^{\prime\prime}}\ ({z_1}^{2} F_{\nu^{\prime\prime}\beta}\ \delta_{\nu^{\prime\prime}\beta})\right] D^{(5)}_{\nu\delta} \\
\hspace{-1.0cm}\ + 16 \left[\sum_{\nu^{\prime}\nu^{\prime\prime}} (D^{(2)}_{\alpha\nu^{\prime}})\ G_{R\nu^{\prime},R\nu^{\prime\prime}} (D^{(2)}_{\nu^{\prime\prime}\beta})\rule{0mm}{6mm}\right] D^{(5)}_{\nu\delta}\\ 
\end{eqnarray*}

Therefore, the sum of all possible scattering diagrams contributing to the four legged vertex (\ shown in the extreme right column of Fig. \ref{fig13}\ ) will be given by
\[
W_{\alpha\beta}^{\gamma\delta}=\sum_{i=A}^{G} \left(W_{\alpha\beta}^{\gamma\delta}\right)_i
\]

\begin{figure}[t]
\centering
\includegraphics[height=8.0cm,width=10cm]{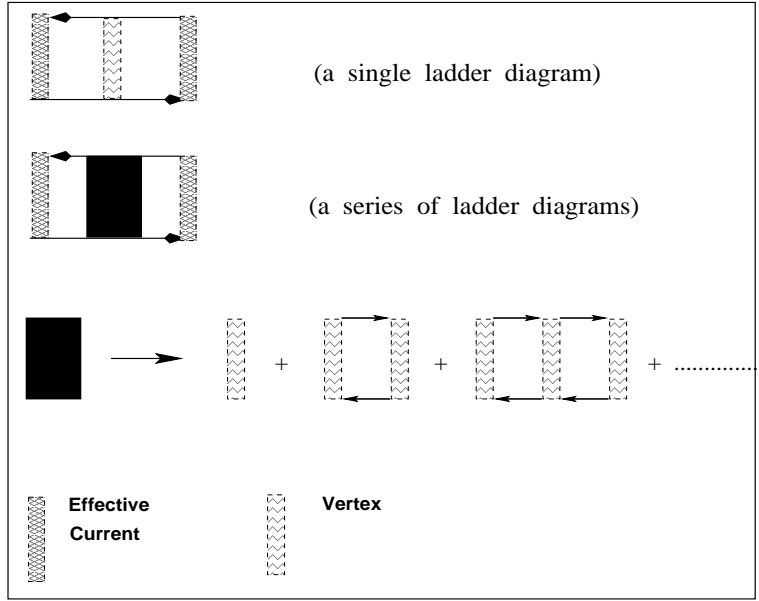}
\caption{The structure of infinite series of ladder diagrams contributing to the correlation function $\ll{\mathbf\kappa}(z_1,z_2)\gg$.}
\label{fig14}
\end{figure}

Here we shall sum the ladder diagrams to all orders. The contribution of a single ladder diagram to the correlation function as shown in the top line of Fig. \ref{fig14} is
\begin{eqnarray}
\hspace{-2.2cm}\sum_{R_1R_2}\sum_{R_3R_4}\sum_{R_5}\sum_{\alpha_1\alpha_2}\sum_{\alpha_3\alpha_4}\sum_{\alpha_5\alpha_6}S_{R_5\alpha_{6},R_1\alpha_1}^{\mathrm {eff}}\ G_{R_1\alpha_{1},R_2\alpha_2}(z_1)\ W_{\alpha_2\alpha_2}^{\alpha_5\alpha_5}\ G_{R_2\alpha_{2},R_3\alpha_3}(z_1)\nonumber\\
\hspace{3.0cm}\left(S_{R_3\alpha_{3},R_4\alpha_4}^{\mathrm {eff}}\right)^{\dagger}\ G_{R_4\alpha_{4},R_2\alpha_5}(z_2)\ G_{R_2\alpha_{5},R_5\alpha_6}(z_2)\
\label{eq24}
\end{eqnarray}

If we apply the homogeneity in full augmented space, it will imply that the above expression is independent of `R' which allows us to take the Fourier transform leading to
\begin{eqnarray}
\hspace{-2.3cm}\left[\rule{0mm}{4mm}\int_{BIZ}\frac{d^3\k}{8\pi^3}\ {\mbf G}(\k,z_2){\S}^{\mathrm{eff}}(\k,z_1,z_2) {\mbf G}(\k,z_1)\right] {\mathbf W}
\left[\rule{0mm}{4mm}\int_{BIZ}\frac{d^3\k^{\prime}}{8\pi^3}\ {\mbf G}(\k^{\prime},z_1)\left({\S}^{\mathrm{eff}}(\k^{\prime},z_1,z_2)\right)^{\dagger} {\mbf G}(\k^{\prime},z_2)\right]\nonumber\\
\hspace{4.5cm}={\mbf \Gamma}(z_1,z_2)\ {\mathbf W}\ \widehat{\mbf \Gamma}(z_1,z_2),\nonumber 
\end{eqnarray}

\n where we have defined 
\begin{eqnarray*}
{\mbf \Gamma}(z_1,z_2) & = &\int_{BIZ}\frac{d^3\k}{8\pi^3}\ {\mbf G}(\k,z_2){\S}^{\mathrm{eff}}(\k,z_1,z_2) {\mbf G}(\k,z_1),\\
\widehat{\mbf \Gamma}(z_1,z_2)&  = &\int_{BIZ}\frac{d^3\k^{\prime}}{8\pi^3}\ {\mbf G}(\k^{\prime},z_1)\left({\S}^{\mathrm{eff}}(\k^{\prime},z_1,z_2)\right)^\dagger\ {\mbf G}(\k^{\prime},z_2).\\
\end{eqnarray*}

Let us now look at the contribution of the infinite series of ladder diagrams (\ shown in the third column of Fig. \ref{fig14}\ ) to the correlation function. Each one of them has the same structure as Equation. (\ref{eq24}), We may then sum up the series as follows :

Let us define  
\begin{eqnarray*}
{\mbf \Theta}_{\alpha\beta}^{\gamma\delta}(z_1,z_2)  =   \int_{BIZ}\frac{d^3\k}{8\pi^3}\ G_{\alpha\beta}(\k,z_1)\ G_{\gamma\delta}(\k,z_2)\nonumber
\end{eqnarray*}

 Then

\[
\fl{\mbf \Lambda}(z_1,z_2) =  {\mbf W} + {\mbf W} {\mbf \Theta} {\mbf W} + {\mbf W} {\mbf \Theta} {\mbf W}  {\mbf \Theta} {\mbf W}+ \ldots 
\ =\ {\mbf W}(z_1,z_2) \left(\rule{0mm}{4mm}{\mbf I}-{\mbf \Theta}(z_1,z_2) {\mbf W}(z_1,z_2)\right)^{-1}. 
\]

Thus the contribution of the infinite series of ladder diagram vertex corrections to the correlation function may be expressed as

\begin{eqnarray}
\hspace{-2cm}\ll \Delta \kappa(z_1,z_2)^{\mathrm ladder}\gg &=& \sum_{\alpha\beta}\sum_{\gamma\delta}{\mbf \Gamma}^{\alpha}_{\beta}(z_1,z_2)\ {\mbf \Lambda}_{\beta\delta}^{\alpha\nu}(z_1,z_2)\ \widehat{\mbf \Gamma}^{\nu}_{\delta}(z_1,z_2) \nonumber\\
&=&  \mathrm{Tr} \left[\rule{0mm}{4mm}{\mbf \Gamma}(z_1,z_2)\otimes \widehat{\mbf \Gamma}(z_1,z_2)\ {\mbf \odot}\ {\mbf \Lambda}(z_1,z_2)\right].
\end{eqnarray}

Including all kinds of disorder corrections, the configuration average of the two-particle Green function (\ or the correlation function\ ) now has the form,
\begin{equation}
\hspace{-2.3cm}\ll \kappa(z_1,z_2) \gg = \ll\kappa_{(1)}(z_1,z_2) \gg + \ll\kappa_{(2)}(z_1,z_2) \gg + \ll \Delta \kappa(z_1,z_2)^{\mathrm ladder}\gg.
\end{equation}

\section{Conclusions.}

We have shown that the effect of disorder scattering is to renormalize the phonon Green functions, via a Dyson equation. Disorder scattering also renormalizes
the heat currents. These corrections belong to two classes : one which is related to the self-energy and another related to the vertex corrections. Once these relations are established, we can use the augmented space recursion algorithm \cite{am2} to obtain the self-energy and the current corrections from it. The expression for the current corrections related to the vertex corrections has also been explicitly shown here. These will form the justification for the calculational algorithm which we had used in our earlier paper on NiPd and NiPt alloys \cite{am}.
\ack One of us (AA) would like to acknowledge financial support from the Council of Scientific and Industrial research (CSIR) India.

\end{document}